\documentclass[12pt,titlepage]{article}
\usepackage{bm, amssymb,pifont,cancel, amsmath,comment,slashed}
\usepackage[dvips]{graphicx}
\usepackage{color}
\makeatletter

\usepackage{ulem}
\usepackage[T1]{fontenc} 
\usepackage{slashed}
\newcommand{\ri}{{\rm i}}

\@addtoreset{equation}{section}
\makeatother
 
\begin{document}

\begin{titlepage}
\null
\begin{flushright}
WU-HEP-24-01
\end{flushright}

\vskip 1.5cm
\begin{center}
\baselineskip 0.8cm
{\LARGE \bf One-loop vacuum energy in 10D super-Yang-Mills theory on magnetized tori with/without 4D $\mathcal{N}=1$ supersymmetric completion}

\lineskip .75em
\vskip 1.5cm

\normalsize

{\large Hiroyuki Abe ${}^a$} $\!${\def\thefootnote{\fnsymbol{footnote}}\footnote[1]{E-mail address: abe@waseda.jp}}, 
{\large Akinari Koichi ${}^a$} $\!${\def\thefootnote{\fnsymbol{footnote}}\footnote[2]{E-mail address: koichi.akinari@suou.waseda.jp}}, 
{\large and} {\large Yusuke Yamada ${}^b$} $\!${\def\thefootnote{\fnsymbol{footnote}}\footnote[3]{E-mail address: y-yamada@aoni.waseda.jp}}

\vskip 1.0em

{ ${}^a$\small\it  Department of Physics, Waseda University, Tokyo 169-8555, Japan}\\
{${}^b$ \small\it  Waseda Institute for Advanced Study, Waseda University, 1-21-1 Nishi Waseda, Shinjuku, Tokyo 169-0051, Japan}

\vspace{13mm}

{\bf Abstract}\\[3mm]
{\parbox{16cm}{\hspace{3mm} \small
We compute the one-loop vacuum energy in 10 dimensional (10D) super Yang-Mills theory compactified on $\mathbb{R}^{1,3}\times (\mathbb{T}^2)^3$ in the presence of the Abelian magnetic fluxes. The regularization of infinite Kaluza-Klein (KK) sum is achieved by the use of Barnes $\zeta$-functions, which can be applied to the case with Landau level structure of KK mass spectrum. We apply the technique to two different models of the 10D super Yang-Mills theory: The one is to introduce the magnetic flux background to the 10D super Yang-Mills action. The other is to first embed 10D super Yang-Mills action into 4D $\mathcal{N}=1$ superspace and to introduce magnetic fluxes. The two model buildings result in tree level potential. The KK mass spectrum of two models look different and we compute the one-loop vacuum energy of each case. Nevertheless, both of the KK mass spectra have the Landau level structure and we can apply the regularization method in the same way. We show that despite the differences the resultant one-loop vacuum energy of both models coincide with each other.
}}

\end{center}

\end{titlepage}

\tableofcontents
\vspace{35pt}
\hrule
\section*{Note added}
After publication of the previous version in PTEP, we have found a significant error in our computation of the one-loop vacuum energy of the model with 4D $\mathcal{N}=1$ supersymmetry completions, which we define below. Accordingly we modified comments and some results from the previous version. We emphasize that most of the results are unchanged and it is yet true that the models with and without 4D ${\cal N}=1$ supersymmetry completions have different tree level potential. Nevertheless, as we will show, the one-loop vacuum energy in both models coincide, which indicates that the KK mass spectra of both models are the same despite differences in their structure. 
\section{Introduction}
Compactified extra dimensions play crucial roles in the construction of realistic models in string theory~\cite{Green:1987mn}. In particular, the chiral gauge theory as well as realistic flavor structure of the standard model have to be realized. A simple realization of such a standard model like spectrum is achieved by extra 6D spaces compactified to $(\mathbb{T}^2)^3$ with magnetic fluxes along tori~\cite{Cremades:2004wa}. The chiral zero-modes realize the standard model like gauge theory, while the degenerate zero modes realize three generations and the overlapping integrals of wave functions on tori can realize the hierarchical Yukawa couplings~\cite{Abe:2012ya,Abe:2012fj}.

Stabilization of moduli fields is an important issue for realistic model buildings. One of the examples is the volume moduli of which vacuum expectation values (VEV) correspond to the physical size of the compactified spaces. Some moduli fields acquire their potential from fluxes of $p$-form fields along compact dimensions~\cite{Gukov:1999ya}, but in general there remain some light moduli. For such light fields, either perturbative~\cite{Berg:2005yu} and/or non-perturbative~\cite{Kachru:2003aw,Conlon:2005ki} corrections are responsible to stabilize them.

KK modes associated with higher-dimensional spaces naturally yield perturbative corrections to moduli potential even though heavy KK modes decouple from low energy effective theory. In particular, their mass spectrum and their symmetry property crucially affect such moduli potential. Technically speaking, the KK mass spectrum depends on the background and we need mathematical methods to regularize the infinite KK sum depending on the KK mass spectrum of each model. 

In this work, we discuss 10D super Yang-Mills theory compactified on tori with Abelian magnetic fluxes where the KK mass spectrum is Landau level form. For such KK mass spectrum, we are able to apply the Barnes $\zeta$-function method to regularize the loop contributions from KK modes. In particular, we apply the method to two different models of magnetized 10D super Yang-Mills models: One is constructed by introducing magnetic fluxes directly to original 10D super Yang-Mills action. Another is first to embed 10D super Yang-Mills theory to 4D $\mathcal{N}=1$ superspace and then to turn on magnetic fluxes. We call the latter to be the case with 4D $\mathcal{N}=1$ supersymmetric completions. Those two model buildings lead to two different 4D effective theories having different tree level flux potential and KK mass spectrum. Despite the difference, the KK mass structures are the same and we are able to apply the Barnes $\zeta$-function regularization technique to both cases. Despite the differences in the KK mass spectrum and tree level potential, we find that the one-loop vacuum energy of the two models coincide with each other, which indicates that the seemingly different KK mass spectra of the two models are the same although it is difficult to prove it by directly comparing the KK mass spectrum of them.

We clarify the reason why we consider the two models: As we show below, the tree level potential of the two models are apparently different, which shows that the model without 4D $\mathcal N=1$ supersymmetry completions does not preserve supersymmetry in the presence of the background magnetic fluxes, independently of the flux configurations. On the other hand, the model with the completions yields tree level potential that vanishes at supersymmetric flux configurations, and the difference of the tree level potential is expected to give other physical differences. Despite such an expectation, our comparison of the one-loop vacuum energy shown below indicates that at least for the free fields coupling to flux backgrounds, there appear no differences between the models with and without 4D $\mathcal{N}=1$ supersymmetry completions except for the tree level flux induced potential. 
 
 We would like to mention related works: In~\cite{Buchmuller:2019zhz}, the authors study the vacuum energy in $SO(32)$ SYM on tori ${(\mathbb T}^2)^3$, and the correspondence between the vacuum energy computed within string theory and field theory has been shown. Furthermore, possible tachyon condensation and supersymmetry restoration have been discussed. We note that the regularization method used there is different from ours. They further study tachyon condensation in type I superstring compactified on a torus in~\cite{Buchmuller:2020nnl}. Those works are mostly based on the 10D super Yang-Mills action with 4D $\mathcal{N}=1$ supersymmetric completions. The one-loop effective potential in 6D super Yang-Mills theory compactified on a single torus with Abelian magnetic fluxes was studied in~\cite{Braun:2006se}. However, in 6D case, magnetic flux potential always breaks supersymmetry unless matter fields get vacuum expectation values (VEV), which is different from the higher dimensional cases. 

 The rest of the paper is organized as follows: In the next section, we review super Yang-Mills theory on tori with Abelian magnetic fluxes. In particular, we use the original 10D action, namely the model without 4D $\mathcal{N}=1$ supersymmetric completions. We review KK mass spectrum within such a model. In Sec.~\ref{SSdescription}, we review the model building based on 4D ${\cal N}=1$ superspace embedding of 10D super Yang-Mills theory with magnetic fluxes, namely the model with 4D $\mathcal{N}=1$ supersymmetric completions. We then compute the one loop vacuum energy. Since the Barnes $\zeta$ function techniques for KK sum regularization are applicable to both cases in the same way, we first consider the case without 4D $\mathcal{N}=1$ supersymmetric completions in Sec.~\ref{vacuumenergy}.  We find that the model without 4D $\mathcal{N}=1$ supersymmetric completions yields UV finite vacuum energy in general, even if the mass spectrum is no longer supersymmetric. 
 We then show the behavior of the finite part of the one loop vacuum energy in a model with and without 4D $\mathcal{N}=1$ supersymmetric completion in Sec.~\ref{example}. It turns out that the one-loop vacuum energy of the two models coincide with each other, namely both give the UV finite one-loop vacuum energy, which we show from numerical results of the finite part whereas the UV finiteness can be confirmed analytically. 
 We then conclude in Sec.~\ref{conclusion}. In appendix~\ref{Barneszeta}, we show the definition of Barnes $\zeta$-function and its analytic continuation. Appendix~\ref{KKexpansionSS} is devoted for the KK expansion within the superspace formulation. We give a few comments related to dimensional reduction in magnetized tori models in appendix~\ref{dimred}.

\section{10D super Yang-Mills model on magnetized tori without 4D $\mathcal{N}=1$ supersymmetric completions}\label{withoutSS}
We review the KK decomposition of ${\mathbb R}^{1,3}\times ({\mathbb T}^2)^n$ ($n=1,2,3$) in the presence of Abelian magnetic fluxes (see~\cite{Cremades:2004wa, Hamada:2012wj} for details\footnote{In~\cite{Conlon:2008qi}, one can find similar computations in different backgrounds.}), and the most of the techniques used here can similarly be applied to the superspace completion model.  In particular, we focus on $U(N)$ super-Yang-Mills action and introduce non-vanishing magnetic fluxes along the compactified dimensions for the diagonal part of vector fields. For simplicity, we first discuss the case $n=1$, namely the six-dimensional case. The metric is given by
\begin{equation}
    ds^2=\eta_{\mu\nu}dx^\mu dx^\nu+g_{\mathtt ij}dx^{\mathtt i}dx^{\mathtt j}=\eta_{\mu\nu}dx^\mu dx^\nu+2(2\pi R_1)^2dz^1d\bar{z}^1
\end{equation}
where $\mu,\nu=0,\cdots,3$, ${\mathtt i}, {\mathtt j}=5,6$, the complex coordinate $z^1$ is related to the real coordinates $x^5$, $x^6$ via $z^1=x^5+\tau x^6$. Here $R_1$ denotes the radius of the torus and $\tau_1$ is the complex structure ($\tau_1\in {\mathbb C}$), and the area of the torus is given by ${\mathcal A}_1=4\pi^2R_1^2{\rm Im}\tau_1$. We impose the periodic boundary condition
\begin{equation}
    z^1\sim z^1+1,\quad z^1\sim z^1+\tau.
\end{equation}

The 6D SYM action without $\mathcal{N}=1$ completion is given by\footnote{We will use bold letters to describe the adjoint ($N \times N$ matrix) representations of the $U(N)$ gauge symmetry.}
\begin{align}
    S=\int d^4x dz^1d\bar{z}^1 \sqrt{g_2}\left[-\frac{1}{4g^2}{\rm Tr}\left({\bm F}_{MN}{\bm F}^{MN}\right)+\frac{\ri}{2g^2}{\rm Tr}(\bar{\bm \lambda} \Gamma^M{\bm D}_M{\bm \lambda})\right],\label{compaction}
\end{align}
where $\sqrt{g_2}$ is the metric determinant on the torus $\mathbb{T}^2$, 
\begin{align}
    {\bm F}_{MN}&=\partial_M{\bm A}_N-\partial_N{\bm A}_M-\ri [{\bm A}_M,{\bm A}_N],\\
    {\bm D}_M{\bm \lambda}&=\partial_M{\bm \lambda} -\ri [{\bm A}_M,{\bm\lambda}],
\end{align}
and the trace ${\rm Tr}$ is taken over the gauge indices. Both the spinors and the vectors are adjoint representations under $U(N)$, and the gauge coupling $g$ has mass dimensions $-1$ in six dimensions. The action can be straightforwardly generalized to $8$ and $10$ dimensional cases. 

We introduce an Abelian magnetic flux along the torus:
\begin{equation}
    {\bm F}=\frac{\ri \pi {\bm M}}{{\rm Im}\tau}dz^1\wedge d\bar{z}^1(={\bm F}_{z^1\bar{z}^1}dz^1\wedge d\bar{z}^1),
\end{equation}
where
\begin{equation}
    {\bm M}=\left(\begin{array}{cccc}
        M^a{\bm 1 }_{N_a\times N_a}&{}&{}&{}\\
        {} & M^b{\bm 1}_{N_b\times N_b}&{}&{}\\
        {}&{}&M^c{\bm 1}_{N_c\times N_c}&{}\\
        {}&{}&{}&\ddots\\
    \end{array}\right),\label{genAflux}
\end{equation}
and $M^{a,b,c}(\in {\mathbb Z})$ are quantized fluxes due to the Dirac quantization condition. Such fluxes break $U(N)$ into $SU(N_a)\times SU(N_b)\times SU(N_c)\times \cdots$, and accordingly the off-diagonal part of adjoint representations couple to the magnetic fluxes. The corresponding background gauge potential is
\begin{equation}
    {\bm A}(z^1,\bar{z}^1)=\frac{\pi {\bm M}}{{\rm Im}\tau}{\rm Im} (\bar{z}^1dz^1),
\end{equation}
from which it follows that 
\begin{align}
    {\bm A}(z^1+1,\bar{z}^1+1)&={\bm A}(z^1,\bar{z}^1)+\frac{\pi {\bm M}}{{\rm Im}\tau}{\rm Im}(dz^1),\\
    {\bm A}(z^1+\tau,\bar{z}^1+\bar\tau)&={\bm A}(z^1,\bar{z}^1)+\frac{\pi {\bm M}}{{\rm Im}\tau}{\rm Im}(\bar\tau dz^1).
    \end{align}
    These shift of the background field can be identified as the gauge transformations with the transformation parameters, respectively,
    \begin{align}
        {\bm \chi}_1=\frac{\pi {\bm M}}{{\rm Im}\tau}{\rm Im}z^1, \quad \chi_2=\frac{\pi \bm{M}}{{\rm Im}\tau}{\rm Im}(\bar\tau z^1).
    \end{align}
One can also introduce Abelian Wilson lines ${\bm \zeta}={\bm\zeta}_1+\tau{\bm\zeta}_2$ where ${\bm \zeta}_{1,2}$ are diagonal real matrices in the gauge basis. Then, the background gauge potential becomes
\begin{align}
    {\bm A}(z,\bar z)=\frac{\pi {\bm M}}{{\rm Im}\tau}{\rm Im}(\bar{z}^1+\bar{\bm\zeta}dz^1).
\end{align}
Note however that the Wilson lines do not change the KK mass spectra of off diagonal modes, and will not contribute to the vacuum energy. Therefore, we may set them to be zero.

In the following, we will derive the KK mass spectrum of fermionic and bosonic fields, respectively. 

\subsection{Spinor}
We will derive the KK mass spectrum of spinor fields. The two dimensional Gamma-matrices on a torus is given by
\begin{align}
    \Gamma^5=\left(\begin{array}{cc}0&1\\ 1&0 \end{array} \right),\quad \Gamma^6=\left(\begin{array}{cc}0&-\ri\\ \ri&0\end{array}\right),
\end{align}
and the fermion wave function on the torus in the spinor basis is represented as 
\begin{align}
    {\bm\psi}=\left(\begin{array}{c}{\bm\psi}_{+}\\ {\bm\psi}_{-}\end{array}\right).
\end{align}
We will implicitly assume that ${\bm\psi}_-={\bm\psi}_+^\dagger$ as ${\bm\psi}$ originates from a Majorana-Weyl spinor in 10D.\footnote{Even in lower dimensions, we would be able to find a representation satisfying ${\bm\psi}_-={\bm\psi}_+^\dagger$ by using symplectic-Majorana Weyl or Majorana conditions in 6D and 8D respectively. Essentially, the spinor is a gaugino and would satisfy the Hermiticity condition. One may also consider dimensional reduction from 10D to lower dimensions.} In particular, here and hereafter we assume that fermions are adjoint representations under $U(N)$. As an illustration, we consider the case for $U(2)$, and ${\bm\psi}_\pm$ in the gauge basis is written as
\begin{align}
    {\bm\psi}_{\pm}=\left(\begin{array}{cc}
        A_{\pm} & B_{\pm}  \\
        C_{\pm} & D_{\pm} 
    \end{array}\right).
\end{align}
Furthermore, consider the magnetic flux of the form
\begin{align}
    {\bm F}=\left(\begin{array}{cc}m^a&0\\ 0&m^b\end{array}\right),
\end{align}
up to an overall factor, where $m^{a,b}\in {\mathbb Z}$.
Then, two dimensional Dirac operator
\begin{align}
    \ri \slashed{\bm D}=\ri \left(\begin{array}{cc}0&-{\bm D}^\dagger\\ {\bm D}&0\end{array}\right)
\end{align}
which act on ${\bm\psi}_\pm$, respectively, as
\begin{align}
    &{\bm D}{\bm\psi}_+=\frac{1}{\pi R_1}\left(\begin{array}{cc}
    \bar{\partial} A_+&\left(\bar\partial +\frac{\pi M^{ab}}{2{\rm Im}\tau_1}(z+\zeta^{ab})\right)B_+\\
   \left(\bar\partial -\frac{\pi M^{ab}}{2{\rm Im}\tau_1}(z+\zeta^{ab})\right)C_+&\bar{\partial}D_+\end{array} \right),\\
    &{\bm D}^\dagger{\bm\psi}_-=\frac{1}{\pi R_1}\left(\begin{array}{cc}
    \partial A_-&\left(\partial -\frac{\pi M^{ab}}{2{\rm Im}\tau_1}(\bar{z}+\bar{\zeta}^{ab})\right)B_-\\
   \left(\partial +\frac{\pi M^{ab}}{2{\rm Im}\tau_1}(\bar{z}+\bar{\zeta}^{ab})\right)C_-&\partial D_- \end{array} \right),
\end{align}
where $M^{ab}=m^a-m^b$ and $\zeta^{ab}=(M^a\zeta^a-M^b\zeta^b)/M^{ab}$.

The zero mode Dirac equation is
\begin{align}
    {\bm D}{\bm\psi}_{+}=0, \quad \bm D^\dagger{\bm\psi}_{-}=0.
\end{align}
We introduce a notation for zero modes of off-diagonal components in gauge basis:
\begin{equation}
    B_+\equiv\psi^{j,M^{ab}}(\tau,z^1+\zeta_{ab}), \quad C_+\equiv\psi^{j,-M^{ab}}(\tau,z^1+\zeta_{ab}),
\end{equation}
where $j$ is the label for degenerate zero mode solutions, and $B_{\pm}=(C_\mp)^*$ due to ${\bm\psi}_-={\bm\psi}_+^\dagger$. It turns out that either $\psi^{j,M^{ab}}$ or $\psi^{j,-M^{ab}}$ can have normalizable solutions, and for $M^{ab}>0$, only $\psi^{j,M^{ab}}$ exists whereas for $M^{ab}<0$, only $\psi^{j,-M^{ab}}$ does. The number of degenerate zero modes is $|M^{ab}|$, namely $j=1,2,\cdots, |M^{ab}|$. 

The differential operators satisfy 
\begin{align}
    (\ri \slashed{\bm D})^2=\left(\begin{array}{cc}{\bm D}^\dagger {\bm D} &0 \\ 0 & {\bm D} {\bm D}^\dagger\end{array}\right)=\frac12\{{\bm D},{\bm D}^\dagger\} +\frac12\left(\begin{array}{cc}[{\bm D}^\dagger, {\bm D}] &0 \\ 0 & -[{\bm D}^\dagger,{\bm D}]\end{array}\right)={\bm \Delta}_2+\left(\begin{array}{cc}\frac{\ri {\bm f}^{\rm eff}_{z\bar{z}}}{(2\pi R)^2} &0 \\ 0 & -\frac{\ri {\bm f}^{\rm eff}_{z\bar{z}}}{(2\pi R)^2}\end{array}\right),
\end{align}
where ${\bm \Delta}_2=\frac12 \{{\bm D},{\bm D}^\dagger\}$ is the two dimensional Laplacian and ${\bm f}^{\rm eff}_{z\bar{z}}$ is the effective field strength for each component of the adjoint representation.\footnote{For instance, $B_+$ have charges $q_a=1, \ q_b=-1$ under the diagonal $U(1)_{a,b}$.} 
Since zero modes $\psi^{j,\pm M^{ab}}(\tau,z^1+\zeta_{ab})$ are annihilated by $D$, acting the square of the Dirac operator reads 
\begin{align}
   \Delta_2\psi^{j,\pm M^{ab}}=\pm \frac{2\pi M^{ab}}{\mathcal{A}_1}\psi^{j,\pm M^{ab}}=\frac{2\pi |M^{ab}|}{\mathcal{A}_1}\psi^{j,\pm M^{ab}},
\end{align}
where the last equality holds as $\psi^{j,\pm M^{ab}}$ exists only when $\pm M^{ab}>0$, respectively. Thus, for off-diagonal elements in gauge basis, the derivative operators satisfy the following algebraic relations
\begin{align}
    N\equiv D^\dagger D=\Delta_2-\frac{2\pi |M^{ab}|}{\mathcal{A}_1},\quad
    [N,D^\dagger]=\frac{4\pi |M^{ab}|}{\mathcal{A}_1}D^\dagger,
\end{align}
 This relation can be regarded as that of the creation and annihilation operators in a harmonic oscillator problem. Indeed, $a\equiv\sqrt{\frac{\mathcal{A}_1}{4\pi |M^{ab}|}}D$, $a^\dagger\equiv\sqrt{\frac{\mathcal{A}_1}{4\pi |M^{ab}|}}D^\dagger$ satisfies the commutation relation of creation and annihilation operators. Then, the zero mode is interpreted as the lowest vacuum state and the $r$-th massive modes can be constructed as
\begin{align}
  \psi^{j,\pm M^{ab}}_r=\frac{1}{\sqrt{r!}}\left(\sqrt{\frac{\mathcal{A}_1}{4\pi |M^{ab}|}}D^\dagger\right)^r\psi^{j,\pm M^{ab}},
\end{align}
and the mass eigenvalues are 
\begin{align}
    m_{F,r,ab}^2=\frac{4\pi |M^{ab}|}{\mathcal{A}_1}r,
\end{align}
where $r=0,1,2,\cdots$. From the relation between the Laplacian and the covariant derivatives, it follows that the eigenvalues of the Laplacian are given by
\begin{align}
    \lambda_r=\frac{2\pi |M^{ab}|}{\mathcal{A}_1}(2r+1),\label{laplacian}
\end{align}
which is useful for later discussions. We note that the mass eigenvalues do not depend on the Wilson line, but the wave function does. In the absence of the flux, mass spectra do depend on the Wilson line. 

The extension to $(\mathbb{T}^2)^n$ can be done straightforwardly by adding quantum numbers from different tori, which yields
\begin{align}
    m_{F,r_1,r_2,r_3,ab}^2=\sum_{n=1}^3\frac{4\pi |M^{ab}_n|}{\mathcal{A}_n}r_n,\label{fermionmass}
\end{align}
and the number of degenerate modes is given by the product of the numbers of flux in each torus, $|M^{ab}_1M_2^{ab}M_3^{ab}|$. Although the above formula is correct, the existence of zero modes is not ensured. A 10D Majorana-Weyl spinor can be decomposed into four 4D Majorana spinors. In general, we are able to consider chirality on each torus, but by fixing the 10D chirality as $\Gamma_{11}\bm\psi=\bm\psi$ where $\Gamma_{11}$ is the 10D analogue of $\gamma_5$ in 4D, the possible helicity assignment of four 4D spinor on three tori is restricted to
\begin{align}
    (+++),(+--),(-+-),(--+),
\end{align}
so that the total helicity is even, which is consistent with the 10D Majorana-Weyl condition. Here, each sign corresponds to the helicity on the 1st, 2nd and 3rd torus. Zero mode exists if and only if the sign of $M_n^{ab}$ matches their helicity on each torus. Therefore, for a fixed $(ab)$, not all of the four 4D fermions have zero modes. Thus, one needs to modify the above mass formula as follows: Noting that $(ba)$-sector feels the magnetic fluxes opposite to that of $(ab)$-sector, we consider both $(ab)$- and $(ba)$-sector simultaneously. Only the signs of $(M^{ab}_1,M_2^{ab},M_3^{ab})$ matter for the existence of zero modes, and there are only two different patterns. One is (1) the case $(M^{ab}_1,M_2^{ab},M_3^{ab})$ all positive and (2) the case where one of $M_1^{ab}$ is positive and others $M_{2,3}^{ab}$ are negative. Other possible assignments would just be derived by changing some labels. Now, for the case (1), the KK towers become
\begin{align}
    (+++):& \ \sum_{n=1}^3\frac{4\pi |M^{ab}_n|}{\mathcal{A}_n}r_n,\label{kk+++}
    \end{align}
    \begin{align}
    (+--):&  \frac{4\pi |M^{ab}_2|}{\mathcal{A}_2}+\frac{4\pi |M^{ab}_3|}{\mathcal{A}_3}+\sum_{n=1}^3\frac{4\pi |M^{ab}_n|}{\mathcal{A}_n}r_n,
    \end{align}
    \begin{align}
    (-+-):& \ \frac{4\pi |M^{ab}_1|}{\mathcal{A}_1}+\frac{4\pi |M^{ab}_3|}{\mathcal{A}_3}+\sum_{n=1}^3\frac{4\pi |M^{ab}_n|}{\mathcal{A}_n}r_n,
    \end{align}
    \begin{align}
    (--+):& \ \frac{4\pi |M^{ab}_1|}{\mathcal{A}_1}+\frac{4\pi |M^{ab}_2|}{\mathcal{A}_2}+\sum_{n=1}^3\frac{4\pi |M^{ab}_n|}{\mathcal{A}_n}r_n,
    \end{align}
    \begin{align}
    (+++)_c:& \ \sum_{n=1}^3\frac{4\pi |M^{ab}_n|}{\mathcal{A}_n}(r_n+1),
    \end{align}
    \begin{align}
    (+--)_c:&  \frac{4\pi |M^{ab}_1|}{\mathcal{A}_1}+\sum_{n=1}^3\frac{4\pi |M^{ab}_n|}{\mathcal{A}_n}r_n,
    \end{align}
    \begin{align}
    (-+-)_c:& \ \frac{4\pi |M^{ab}_2|}{\mathcal{A}_2}+\sum_{n=1}^3\frac{4\pi |M^{ab}_n|}{\mathcal{A}_n}r_n,\end{align}
    \begin{align}
    (--+)_c:& \ \frac{4\pi |M^{ab}_3|}{\mathcal{A}_3}+\sum_{n=1}^3\frac{4\pi |M^{ab}_n|}{\mathcal{A}_n}r_n,
\end{align}
where $r_n=0,1,\cdots$ and $(\cdot\cdot\cdot)_c$ denotes the $(ba)$-sector feeling magnetic fluxes opposite to $M^{ab}_n$. Thus, we see that the zero mode appears in $(+++)$. For the case (2), the flux difference has the sign $(+,-,-)$ and the fermionic KK masses are
\begin{align}
    (+++):& \ \frac{4\pi |M^{ab}_2|}{\mathcal{A}_2}+\frac{4\pi |M^{ab}_3|}{\mathcal{A}_3}+\sum_{n=1}^3\frac{4\pi |M^{ab}_n|}{\mathcal{A}_n}r_n,\end{align}
    \begin{align}
    (+--):&  \sum_{n=1}^3\frac{4\pi |M^{ab}_n|}{\mathcal{A}_n}r_n,\end{align}
    \begin{align}
    (-+-):& \ \frac{4\pi |M^{ab}_1|}{\mathcal{A}_1}+\frac{4\pi |M^{ab}_2|}{\mathcal{A}_2}+\sum_{n=1}^3\frac{4\pi |M^{ab}_n|}{\mathcal{A}_n}r_n,\end{align}
    \begin{align}
    (--+):& \ \frac{4\pi |M^{ab}_1|}{\mathcal{A}_1}+\frac{4\pi |M^{ab}_3|}{\mathcal{A}_3}+\sum_{n=1}^3\frac{4\pi |M^{ab}_n|}{\mathcal{A}_n}r_n,\end{align}
    \begin{align}
    (+++)_c:& \ \frac{4\pi |M^{ab}_1|}{\mathcal{A}_1}+ \sum_{n=1}^3\frac{4\pi |M^{ab}_n|}{\mathcal{A}_n}(r_n+1),\end{align}
    \begin{align}
    (+--)_c:& \sum_{n=1}^3\frac{4\pi |M^{ab}_n|}{\mathcal{A}_n}(r_n+1),\end{align}
    \begin{align}
    (-+-)_c:& \ \frac{4\pi |M^{ab}_3|}{\mathcal{A}_3}+\sum_{n=1}^3\frac{4\pi |M^{ab}_n|}{\mathcal{A}_n}r_n,\end{align}
    \begin{align}
    (--+)_c:& \ \frac{4\pi |M^{ab}_2|}{\mathcal{A}_2}+\sum_{n=1}^3\frac{4\pi |M^{ab}_n|}{\mathcal{A}_n}r_n,\label{kk--+}
\end{align}
where $r_n=0,1,\cdots$. We see that although KK mass towers of each mode in the case (1) and (2) differ from each other, the eigenvalues appearing in the spectra are the same. This means that the chirality assignments do not affect the vacuum energy, and therefore we do not need to specify the chirality assignments and both cases are treated simultaneously. 

One also needs to consider a larger gauge group $U(N)$ $(N>5)$ for realistic models including the standard model.\footnote{Semi-realistic models have been constructed e.g. in~\cite{Abe:2012ya,Abe:2012fj} with $U(8)$ gauge theory.} In such a case, we need to consider fluxes of the form~\eqref{genAflux}. Even in such a case, the formulas shown so far formally do not change except that the off-diagonal sector $\psi^{ab}$ becomes bifundamental representations under $SU(N_a)$ and $SU(N_b)$, and therefore, the number of zero modes should be counted as $|M_{ab}|N_aN_b$, which is important for the later computation of vacuum energy. This argument applies to the bosonic modes discussed below. 

\subsection{Bosonic fields}
Here we will derive the KK mass spectrum of bosonic fields. For an adjoint scalar field, the mode functions $\phi_r^{ab}(z,\bar{z})$ are given by the 2D Laplace equation
\begin{align}
\Delta_2\phi^{ab}_r(z,\bar{z})=m_r^2\phi^{ab}_r(z,\bar{z}),
\end{align}
and since we know the Laplacian eigenvalues~\eqref{laplacian}, the KK mass $m_{S,r,ab}^2$ is found to be
\begin{align}
    m_{S,ab,r}^2=\frac{2\pi |M^{ab}|}{\mathcal{A}_1}(2r+1).\label{scalarmass}
\end{align}
In type IIB string models, such charged scalars may appear in the effective action of e.g. D$p$-branes with $p=5,7$, but not from 10D super-Yang-Mills theory/D$9$-brane models, where no fundamental charged scalars exist.

Next, we discuss the mass spectrum of vector fields. Since we are concerned with the effective theory of 10D SYM, there are no fundamental charged vector fields, and such fields appear only as the off-diagonal part of the adjoint representations in a large gauge group, which is broken to smaller ones due to the background magnetic fields. In order to separate the diagonal and the off-diagonal part of the adjoint representations, we use the basis of $U(N)$ given by diagonal generators ${\bm U}^a$ and off-diagonal ones ${\bm e}_{ab}$ where $({\bm U}^a)^i_j=\delta_{ai}\delta_{aj}$, and $({\bm e}_{ab})^i_j=\delta_{ai}\delta_{bj}$ ($a,b,i,j=1,\cdots,N$ and $a\neq b$). With this basis, a $U(N)$ vector potential is decomposed as
\begin{align}
    {\bm A}_M=B_M^a{\bm U}^a+W_M^{ab}{\bm e}_{ab},
\end{align}
where $W^{ab}_M=(W^{ba}_M)^*$.  The quadratic part of the action is 
\begin{align}
    -\frac{1}{4g^2}{\rm Tr}({\bm F}^{MN}{\bm F}_{MN})=&-\frac{1}{4g^2}G^a_{MN}G^{aMN}+\frac{\ri}{4g^2}(G^{aMN}-G^{bMN})(W_M^{ab}W^{ba}_N-W_N^{ab}W_M^{ba})\nonumber\\
    &-\frac{1}{2g^2}(D_MW_N^{ab}D^MW^{baN}-D_NW_M^{ab}D^MW^{baN})+\cdots\label{FFdecomp}
\end{align}
 where the ellipses denote terms cubic or higher in quantum fields and 
\begin{align}
    G^{a}_{MN}&=\partial_MB_N^a-\partial_NB^a_M,\\
    D_MW_N^{ab}&=\partial_MW_N^{ab}-\ri(B^a_M-B_M^b)W_N^{ab}.
\end{align}
We consider only (constant) Abelian magnetic fluxes, namely, the background field value is given only to $B^a_M(z)$ or equivalently $G_{MN}^a(={\rm const.})$. We also notice that the Abelian vector potentials $B^a$ themselves do not couple to the magnetic flux and their KK masses are given by the eigenvalues of the Laplacian on tori without magnetic fields. The E.O.M of charged vector bosons are
\begin{align}
   &\Box W_{\mu}^{ab}-\Delta_2 W_{\mu}^{ab}=0\\
   &\Box W_{\mathtt m}^{ab}-\Delta_2 W_{\mathtt m}^{ab}+2\ri \langle G^{ab}_{\mathtt {nm}}\rangle W^{{\mathtt n}ab}=0,\label{vectorEOM}
\end{align}
where ${\mathtt m}, {\mathtt n}=z,\bar{z}$, $G_{\mathtt{mn}}^{ab}=G^{a}_{\mathtt{mn}}-G_{\mathtt{mn}}^b$, we have taken the gauge $\partial_\mu W^{\mu ab}=0=D^{\mathtt m}W_{\mathtt m}^{ab}$, and $\Delta_2=-D_{\mathtt m}D^{\mathtt m}$ denotes the covariant Laplacian on $\mathbb{T}^2$. In the above equations, background fields are included only in the covariant derivatives. As shown in \eqref{vectorEOM}, the vector fields $W_{\mathtt m}^{ab}$ effectively couple to the relative flux $G_{\mathtt{mn}}^{ab}$. The KK masses of 4D components $W_\mu^{ab}$, $m_{V,r,ab}^2$, are the same as that of scalars~\eqref{scalarmass}. Expanding the charged vector fields $W_{\mathtt m}^{ab}$ as
\begin{align}
    W_{\mathtt m}^{ab}(x,y)=\sum_{r}\varphi_{{\mathtt m},r}^{ab}(x)\phi_{{\mathtt m},r}^{ab}(y),
\end{align}
the E.O.M reads (with the complex basis)
\begin{align}
    \left(\Delta_2-\frac{4\pi M^{ab}}{\mathcal{A}_1}\right)\phi_{z,r}^{ab}=&m_{z,ab,r}^2\phi_{z,r}^{ab},\\
    \left(\Delta_2+\frac{4\pi M^{ab}}{\mathcal{A}_1}\right)\phi_{\bar{z},r}^{ab}=&m_{\bar{z},ab,r}^2\phi^{ab}_{\bar{z},r},
\end{align}
where we have defined the KK mass eigenvalues by $\Box\varphi^{ab}_{z ,r}=m_{z,r,ab}^2\varphi^{ab}_{z,r}$.
Thus, using \eqref{scalarmass}, we find the $r$-th KK mass eigenvalues of $A_{z(\bar{z})}$ to be
\begin{equation}
    \frac{2\pi|M^{ab}|}{\mathcal{A}_1}(2r+1)\mp \frac{4\pi M^{ab}}{\mathcal{A}_1},
\end{equation}
respectively. Notice that both for $M^{ab}>0$ and for $M^{ab}<0$, there appears a tachyonic mode for $r=0$. Such a tachyonic mode may disappear only when we consider the case $({\mathbb T}^2)^n$ ($n\geq2$) : Notice that it is straightforward to generalize \eqref{vectorEOM} to the 10D cases by replacing $\Delta_2$ with the Laplacian $\Delta_{2n}$ on $({\mathbb T}^2)^n$, which yields e.g. the mass eigenvalues of $A_{z^1},A_{\bar{z}^1}$ as
\begin{align}
m^2_{z^1(\bar{z}^1), ab,(r_1,r_2,r_3)}=2\pi\left(\frac{|M_1^{ab}|}{\mathcal{A}_1}(2r_1+1)\mp\frac{2M_1^{ab}}{\mathcal{A}_1}+\frac{|M_2^{ab}|}{\mathcal{A}_2}(2r_2+1)+\frac{|M_3^{ab}|}{\mathcal{A}_3}(2r_3+1)\right),\label{vectormass}
\end{align}
where $M^{ab}_{n}$ denotes the relative flux on the $n$-th torus. Now the lowest eigenvalue is
\begin{equation}
    2\pi\left(\frac{|M_2^{ab}|}{\mathcal{A}_2}+\frac{|M_3^{ab}|}{\mathcal{A}_3}-\frac{|M_1^{ab}|}{\mathcal{A}_1}\right),\label{vectorlowest}
\end{equation}
which can be zero or positive for suitably chosen parameters. The KK masses of $A_{z^{2,3}},A_{\bar{z}^{2,3}}$ are derived in the same way.
\subsection{Supersymmetric mass spectrum}\label{SSmasswo}
We show that the KK mass spectrum derived so far can be supersymmetric with a particular set of fluxes and areas of tori. First, we recall that the fermionic KK mass patterns are independent of the flux signs if we take account all the possible chirality modes. The same is true for bosonic fields since 4D vector masses are the same as scalars in \eqref{scalarmass} which is independent of flux signs and the extra dimensional vector fields $\varphi_{z_i,r_i}^{ab}$ have the mass pattern shown in \eqref{vectormass}, which shows that the flip of the flux signs does not change the total spectrum. Therefore, we can choose all the flux to be positive without loss of generality.

As fermionic fields contain a zero mode, supersymmetry would be realized only when there appears a bosonic zero mode. Indeed, it is true: Due to the symmetry of the mass formula, we may assume that $\varphi_{z_1,0,0,0}^{ab}$ to be a zero mode, which implies that the mass~\eqref{vectorlowest} vanishes, or equivalently,
\begin{align}
    \frac{M_2^{ab}}{\mathcal{A}_2}+\frac{M_3^{ab}}{\mathcal{A}_3}=\frac{M_1^{ab}}{\mathcal{A}_1}.\label{masscond}
\end{align}
In this case, the mass formula~\eqref{vectormass} can be rewritten as
\begin{align}
    m^2_{z^1, ab,(r_1,r_2,r_3)}=&\sum_{i=1}^34\pi\frac{M_i^{ab}}{\mathcal{A}_i}r_i,\\
    m^2_{\bar{z}^1,ab,(r_1,r_2,r_3)}=&8\pi\frac{M_2^{ab}}{\mathcal{A}_2}+8\pi\frac{M_3^{ab}}{\mathcal{A}_3}+\sum_{i=1}^34\pi\frac{M_i^{ab}}{\mathcal{A}_i}r_i.
\end{align}
Similarly
\begin{align}
    m^2_{z^2,ab,(r_1,r_2,r_3)}=&4\pi\frac{M_3^{ab}}{\mathcal{A}_3}+\sum_{i=1}^34\pi\frac{M_i^{ab}}{\mathcal{A}_i}r_i,\end{align}
    \begin{align}
    m^2_{\bar{z}^2,ab,(r_1,r_2,r_3)}=&8\pi\frac{M_2^{ab}}{\mathcal{A}_2}+4\pi\frac{M_3^{ab}}{\mathcal{A}_3}+\sum_{i=1}^34\pi\frac{M_i^{ab}}{\mathcal{A}_i}r_i,\end{align}
    \begin{align}
     m^2_{z^3, ab,(r_1,r_2,r_3)}=&4\pi\frac{M_2^{ab}}{\mathcal{A}_2}+\sum_{i=1}^34\pi\frac{M_i^{ab}}{\mathcal{A}_i}r_i,\end{align}
    \begin{align}
    m^2_{\bar{z}^3,ab,(r_1,r_2,r_3)}=&4\pi\frac{M_2^{ab}}{\mathcal{A}_2}+4\pi\frac{M_3^{ab}}{\mathcal{A}_3}+\sum_{i=1}^34\pi\frac{M_i^{ab}}{\mathcal{A}_i}r_i,\end{align}
    \begin{align}
    m^2_{V,ab,(r_1,r_2,r_3)}=&4\pi\frac{M_2^{ab}}{\mathcal{A}_2}+4\pi\frac{M_3^{ab}}{\mathcal{A}_3}+\sum_{i=1}^34\pi\frac{M_i^{ab}}{\mathcal{A}_i}r_i,
\end{align}
where we have used \eqref{masscond}. One may substitute \eqref{masscond} to the fermionic mass formulas and immediately find that there appear the same set of KK towers. Thus, we find that the KK mass spectrum can be supersymmetric. 

As we will show explicitly, the models with 4D $\mathcal{N}=1$ supersymmetric completions can be supersymmetric if there appears a bosonic massless mode as is the above case. Nevertheless, we will also find that the supersymmetric KK mass spectrum patterns differ from the above ones.

\section{10D super Yang-Mills model on magnetized tori with 4D $\mathcal{N}=1$ supersymmetric completions}\label{SSdescription}
 We briefly review 4D $\mathcal{N}=1$ superfield description of 10D super-Yang-Mills theory on magnetized tori on the basis of~\cite{Abe:2012ya}. The 10D super Yang-Mills fields are repackaged into 4D $\mathcal{N}=1$ superfields as
 \begin{align}
     {\bm V}(X,\theta,\bar\theta)=&\cdots -\theta\sigma^\mu\bar\theta {\bm A}_\mu(X)+\ri\bar\theta^2\theta{\bm \lambda}_0-\ri\theta^2\bar\theta\bar{\bm\lambda}_0+\frac12 {\bm D},\\
     {\bm\phi}_i(Y,\theta)=&\frac{1}{\sqrt2}{\bm A}_i(Y)+\sqrt{2}\theta{\bm\lambda}_i(Y)+\theta{\bm F}_i(Y),
 \end{align}
 where $X^M$ denotes ten-dimensional spacetime coordinates and $Y^M=(y^\mu,x^{m})$, $(\mu=0,\cdots,3, m=4,\cdots,9)$ with $y^\mu=x^\mu+\ri \theta\sigma^\mu\bar\theta$, and ${\bm\lambda}_0={\bm\lambda}_{+++}$, ${\bm\lambda}_1={\bm\lambda}_{+--}$, ${\bm\lambda}_2={\bm\lambda}_{-+-}$, ${\bm\lambda}_{3}={\bm\lambda}_{--+}$. Note that the chirality labels on the r.h.s. are the same as one we use in the previous section. For the vector superfield ${\bm V}$, the ellipses denote the components vanishing in the Wess-Zumino gauge. The complexified vector fields along compact dimensions are defined as
 \begin{align}
     {\bm A}_i=-\frac{1}{{\rm Im}\tau_i}(\tau_i^*A_{2+2i}-A_{3+2i}),
 \end{align}
 where $i=1,2,3$. The superfields contain $F$- and $D$-terms absent in our component action. With the superfield, 10D magnetized super-Yang-Mills action can be written as
 \begin{align}
     S=\int d^{10}X\sqrt{-G}\left[\int d^4\theta K+\left\{\int d^2\theta\left(\frac{1}{4g^2}{\rm Tr}{\bm W}^\alpha{\bm W}_\alpha+W\right)+{\rm h.c.}\right\}\right],
 \end{align}
 where 
 \begin{align}
     K=&\frac{2}{g^2}h^{\bar{i}j}{\rm Tr}\left[(\sqrt{2}\bar{\partial}_{\bar i}+\bar{\bm\phi}_{\bar i})e^{-{\bm V}}(-\sqrt2\partial_j+{\bm \phi}_j)e^{\bm V}+\bar{\partial}_{\bar i}e^{-\bm V}\partial_j e^{\bm V}\right],\\
    {\bm W}_\alpha=&-\frac14\bar{D}^2e^{-\bm V}D_\alpha e^{\bm V},\\
    W=&\frac{1}{g^2}\epsilon^{\rm ijk}e_{\rm i}{}^ie_{\rm j}{}^je_{\rm k}{}^k{\rm Tr}\left[\sqrt{2}{\bm \phi}_i\left(\partial_j{\bm \phi}_k-\frac{1}{3\sqrt2}[{\bm \phi}_j,{\bm\phi}_k]\right)\right],
 \end{align}
the totally anti-symmetric tensor $\epsilon^{\rm ijk}$ satisfies $\epsilon^{123}=1$ and the complexfied metric $h_{i\bar{j}}$ and its vielbein $e_{\rm i}{}^i$ are\footnote{The complexified metric in~\cite{Abe:2012ya,Abe:2012fj} is incorrect and must be replaced as $h_{i\bar{j}}=2(2\pi R_i)\delta_{i\bar j}\to2(\pi R_i)\delta_{i\bar j}$. } 
\begin{align}
h_{i\bar j}=&2(\pi R_i)^2\delta_{i\bar{j}}=\delta_{\rm i\bar{\rm j}}e_i{}^{\rm i}\delta_{\rm i\bar{\rm j}}e_{\bar j}{}^{\bar{\rm j}},\\
e_i{}^{\rm i}=&\sqrt{2}(\pi R_i)\delta_i{}^{\rm i}.
\end{align}
Note that the complexified metric is given by
\begin{align}
    ds_{6D}^2=2h_{i\bar{j}}dz^id\bar{z}^{\bar j}.
\end{align}
The Roman indices ($\rm i,j,k$) are the local Lorentz frame coordinates. In the Wess-Zumino gauge, ${\bm V}^3=0$, the K\"ahler potential can be simply written as
\begin{align}
    K=\frac{2}{g^2}h^{{\bar i}j}{\rm Tr}\left[\bar{\bm\phi}_{\bar i}{\bm\phi}_j-\sqrt{2}(\bar{\bm D}^\dagger_j\bar{\bm\phi}_{\bar i}+{\bm D}^\dagger_{\bar i}{\bm \phi}_j){\bm V}+[\bar{\bm\phi}_{\bar i},{\bm\phi}_j]{\bm V}-{\bm D}_{\bar i}^\dagger{\bm V}{\bm D}_j{\bm V}\right],
\end{align}
where we have performed integration by parts with respect to derivatives on compact dimensions and the covariant derivatives are defined as ${\bm D}_j=\partial_j-\frac{1}{\sqrt2}[{\bm\phi}_j,\cdot\ ]$ and ${\bm D}^\dagger_{\bar i}=-(\bar{\partial}_{\bar i}+\frac{1}{\sqrt2}[\bar{\bm \phi}_{\bar i},\cdot\ ])$. With the covariant derivative, we may write the superpotential as
\begin{align}
    W=&\frac{1}{g^2}\epsilon^{\rm ijk}e_{\rm i}{}^ie_{\rm j}{}^je_{\rm k}{}^k{\rm Tr}\left[\sqrt{2}{\bm \phi}_i\left({\bm D}_j{\bm \phi}_k+\frac{2}{3\sqrt2}[{\bm \phi}_j,{\bm\phi}_k]\right)\right]\label{10Dsuperpotential}
\end{align}

We introduce constant Abelian magnetic fluxes $\langle{\bm \phi}\rangle={\bm B}_i/\sqrt2$ and make a replacement ${\bm\phi}_i\to {\bm \phi}_i+{\bm B}_i/\sqrt2$. In particular we assume $\partial_j{\bm B}_i=0$ for $i\neq j$, which means the absence of oblique fluxes. The K\"ahler and super-potential with such a background are
\begin{align}
    K=&\frac{1}{g^2}\sum_{i=1}^3\frac{1}{(\pi R_i)^2}\Biggl[\bar{\phi}^{ab}_{\bar i}{\phi}^{ab}_i+2f_{\bar{i}i}^{a}V^{aa}-\sqrt{2}(\bar{ D}^{\dagger ba}_i\bar{\phi}^{ba}_{\bar i}+{D}^{\dagger ab}_{\bar i}{\phi}^{ab}_i){V}^{ba}+([\bar{\bm\phi}_{\bar i},{\bm\phi}_i])^{ab}{V}^{ba}\nonumber\\
    &-\left({D}^{\dagger ab}_{\bar i}{V}^{ab}-\frac{1}{\sqrt2}([\bar{\bm\phi}_{\bar i},{\bm V}])^{ab}\right)\left(D^{ba}_i{V}^{ba}-\frac{1}{\sqrt2}([{\bm\phi}_i,{\bm V}])^{ba}\right)\Biggr],\label{10DKahler}\\
    W=&\frac{1}{g^2}\frac{1}{2 \prod_{l=1}^3(\pi R_l)}\epsilon^{\rm ijk}\delta_{\rm i}{}^i\delta_{\rm j}{}^j\delta_{\rm k}{}^k\left[\phi_{i}^{ab}D_j^{ba}\phi_k^{ba}-\frac{1}{3\sqrt2}\phi_i^{ab}([{\bm\phi_j,\bm\phi_k}])^{ba}\right],
\end{align}
where $a,b$ are gauge indices, $f_{\bar{i}j}^a=\partial_j{B}^{a}_{\bar i}=\bar{\partial}_{\bar{i}}B^{a}_j$ and the covariant derivatives contain only the background fields. Here and hereafter we take the background fields to be
\begin{align}
    B^{a}_i=\frac{\pi}{{\rm Im}\tau_i}M^{a}_i\bar{z}^{i}.
\end{align}
The flux parameter $M^{a}_i$ introduced above it is the same as ours. Accordingly the covariant derivatives become $D^{ab}_i=(\partial_i-\frac{\pi}{2{\rm Im}\tau_i}M^{ab}_i\bar{z}^i)$, $\bar{D}_{\bar i}^{ab}=-(\bar{\partial}_{\bar i}+\frac{\pi}{2{\rm Im}\tau_i}M^{ab}_i z^i)$ where $M_i^{ab}=M_i^a-M^b_i$. 

The procedure to perform KK expansion is quite similar to the case without 4D $\mathcal{N}=1$ supersymmetric completions since the property of the mode functions are the same. However, in superspace approach, we may directly use superfields for KK expansion and discuss the KK expansion of each superfield in Appendix~\ref{KKexpansionSS} with a specific background since the procedure is more involved. As we show there, all mass terms but flux induced D-term scalar masses can be written at the level of superfields as explicitly shown in Appendix~\ref{KKexpansionSS}. Namely, vanishing flux-induced D-terms immediately leads to manifest 4D $\mathcal{N}=1$ supersymmetry.

\subsection{Note on the differences between models with and without 4D $\mathcal{N}=1$ supersymmetric completions}\label{diffSS}
Now, let us see the difference of the models with and without 4D $\mathcal{N}=1$ supersymmetric completions. We will focus on the 4D effective potential associated with the magnetic fluxes. In the model without 4D $\mathcal{N}=1$ supersymmetric completions, we find
\begin{align}
    \mathcal{L}_G=-\frac{1}{4g^2}\sum_{i=1}^3\sum_{a}G_{z^i\bar{z}^i}^aG^{az^i\bar{z}^i}=-\frac{1}{4g^2}\sum_{i=1}^3\sum_{a}\frac{\pi^2 (M^{a}_i)^2}{(4\pi^2 R_i^2{\rm Im}\tau_i)^2},\label{fluxenergycomp}
\end{align}
Obviously, the energy is positive definite and non-vanishing unless all the magnetic fluxes are turned off, which shows the spontaneous breaking of all supersymmetries.

Let us then consider the flux induced potential in the model with 4D $\mathcal{N}=1$ supersymmetic completions, which is given by D-term potential. In \eqref{10DKahler}, assuming that fluctuation superfields $\phi_i$ have vanishing VEV, we find tadpoles of $V^{aa}$ as 
\begin{align}
    K\supset \frac{1}{g^2}\sum_{i=1}^3\frac{2}{\pi R_i}f_{\bar{i}i}^aV^{aa}.
\end{align}
Note that since diagonal elements feel no magnetic flux, their wave functions are trivial, and therefore, in performing integration over compact spaces, only the zero modes of $V^{aa}$ obtain these tadpoles. For a fixed $a$, the tadpole is proportional to
\begin{align}
   D\propto \sum_{i=1}^3\frac{2}{\pi R_i}f_{\bar{i}i}^a,
\end{align}
which would give rise to potential
\begin{align}
    V_D\propto \left(\sum_{i=1}^3\frac{2}{\pi R_i}f_{\bar{i}i}^a\right)^2.\label{dtermflux}
\end{align}
Notice the significant difference from the flux potential shown in \eqref{fluxenergycomp}, which is sum of square whereas the above D-term potential is square of sum! Therefore, in the model with 4D $\mathcal{N}=1$ supersymmetric completions, flux potential energy may vanish even for non-zero fluxes when their values are appropriately chosen. Note that the above D-term tadpole is incomplete as the integration over tori is not yet achieved in the above expression, but it does not change the above discussion much.

What would be the origin of the difference in the tree level flux induced potential? To clarify the difference, let us consider the following seemingly total derivative terms that indeed become surface terms in trivial backgrounds: 
\begin{align}
    \mathcal{L}_{G}'=-\frac{1}{16g^2}\sum_{i=1}^3a_i\epsilon^{m_in_ip_iq_i}G^a_{m_in_i}G^a_{p_iq_i},\label{LGp}
\end{align}
where $\epsilon^{m_in_ip_iq_i}$ is a completely anti-symmetric tensor, $(m_i,n_i,p_i,q_i)$ runs over $(7,8,9,10)$, $(9,10,5,6)$, $(5,6,7,8)$  for $i=1,2,3$ respectively.\footnote{Namely $(m_i,n_i,p_i,q_i)$ do not contain the coordinates of the $i$-th torus.} In the absence of the oblique fluxes, namely the fluxes along the different tori such as $G_{57}$ or $G_{59}$, the total derivative terms become
\begin{align}
    \mathcal{L}_{G}'=-\frac{1}{2g^2}\sum_a (a_1G^a_{78}G^a_{9,10}+a_2G^a_{56}G^a_{9,10}+a_3G^a_{56}G^a_{78}).\label{totdivterms}
\end{align}
For instance, taking $a_1=a_2=a_3=1$ and adding the terms to the kinetic term, we find the total action to be
\begin{align}
    \mathcal{L}_G^{\rm tot}=-\frac{1}{4g^2}(G^a_{56}+G^a_{78}+G^a_{9,10})^2,\label{apluspot}
\end{align}
where we have used real coordinates and took $\tau_i=\ri$ for simplicity. Note that we could take $a_1=-a_2=-a_3=1$, which yields
\begin{align}
    \mathcal{L}_G^{\rm tot}=-\frac{1}{4g^2}(-G^a_{56}+G^a_{78}+G^a_{9,10})^2.
\end{align}
The resultant potential energy~\eqref{apluspot} has the structure of the D-term potential in \eqref{dtermflux} (after appropriately compensated by metric). Note however, addition of the above terms seems to break Lorentz symmetry in the original 10D theory since the added terms have indices on two tori but not all. Such a violation of manifest 10D Lorentz invariance is necessary to rewrite higher dimensional supersymmetric action in terms of 4D $\mathcal{N}=1$ superfields. In the presence of nontrivial backgrounds, the violation of the 10D Lorentz invariance becomes not just a technical problem but a physical consequence of the backgrounds. The absence of Lorentz symmetry in 10D would explicitly break extended supersymmetry in 10D super Yang-Mills theory. 

As we have discussed above, there are some choices of parameters $a_i$ above, and it seems consistent with the number of supersymmetry: 10D supersymmetry or equivalently $\mathcal{N}=4$ 4D supersymmetry has four 4D $\mathcal{N}=1$ supersymmetries, whereas possible patterns of $a_i$ are $a_i=\pm1$ or one of them has the opposite sign, and the total number of the choices is four, the same number as that of 4D $\mathcal{N}=1$ supersymmetries. We expect that each choice of $\{a_i\}$ leaves one of 4D $\mathcal{N}=1$ supersymmetries among four of them. In the commonly used formulation, $a_i=1$ is used, which manifests SU(3) symmetry of superfields ${\bm\phi}_i$~\cite{Arkani-Hamed:2001vvu}.

We would like to give a few comments on the relations to superstring theory. If magnetic fluxes are introduced on D9-branes, which are possible origins of our magnetized models, there would be positive energy due to branes. Forming D-term potential in the complete square form requires products of fluxes from different tori, which can be either positive or negative. Such contributions may originate from the tension of D5-brane or O5-plane, depending on its sign, as in a supersymmetric D9/D5 system. As a result, magnetic flux energy can be canceled within string theory. Although our field theoretical models cannot be directly related to string theoretical models, we suspect that the superspace completion may play similar roles to e.g. O5 plane, as the compensating flux terms can be written as 6 dimensional object.\footnote{Recall that the added terms have Lorentz indices on two tori among three, and integration over these tori leave 6D volume with products of fluxes from each torus, which implies 6D object. See ~\cite{Angelantonj:2000hi} for a related discussion.} It would be interesting to study the relation to D-branes. One of the ways to address such an issue would be to consider the DBI action of non-Abelian gauge theory along the line of the study in~\cite{Abe:2021uxb}. 

We have focused only on the bosonic part of the 4D $\mathcal{N=1}$ supersymmetric completions but there can be associated fermionic terms in general. Nevertheless, we will show that the one-loop vacuum energy of the models with and without 4D $\mathcal{N=1}$ supersymmetric completions are the same, which indicates that the KK spectra of the two models coincide. We will not prove the correspondence by a direct comparison of the two KK spectra, but just indirectly show it via the behavior of the one-loop vacuum energy.

We finally give a few comments on oblique fluxes. In \eqref{totdivterms}, we have shown the products of field strength on the same torus since we have assumed the absence of oblique fluxes. However, if they exist, they give nontrivial contributions to \eqref{LGp}, which are repackaged into superpotential~\eqref{10Dsuperpotential}. Therefore, 4D $\mathcal{N}=1$ supersymmetric completions of any types of magnetic flux backgrounds require the terms~\eqref{LGp}.

\section{The one-loop vacuum energy in magnetized tori model without 4D $\mathcal{N}=1$ supersymmetric completions}\label{vacuumenergy}
In this section, we compute the vacuum energy associated with bulk fields charged under the Abelian subgroup having magnetic fluxes. Since the computational method is completely the same in the models with and without the 4D $\mathcal{N}=1$ supersymmetric completions, we first discuss the former case.

The basic formula in computing the vacuum energy is
\begin{align}
    V=&\sum_{s,\vec{r}}(-1)^{2s}\mu^{4-d}\int \frac{d^dk}{(2\pi)^d}\log\left(\frac{k^2+m_{s,\vec{r}}^2}{\mu^2}\right)\nonumber\\
    =&-\frac{\Gamma\left(-\frac{d}{2}\right)}{(4\pi)^{\frac d2}}\sum_{s,\vec{r}}(-1)^{2s}(m_{s,\vec{r}}^2)^{\frac d2}\mu^{4-d},\label{vacuumformula}
\end{align}
 where $s$ represents a formal label of spin and representation under $U(N)$ and $\vec{r}$ formal KK numbers and $\mu$ denotes an arbitrary renormalization scale parameter. The formal factor $(-1)^{2s}$ becomes $-1$ for fermionic modes and $1$ otherwise. The KK mass spectrum discussed in the previous section gives us the explicit forms of $m_{s,\vec r}^2$ for all~$s$. The above expression is not yet completely regularized because of the infinite sum of KK modes. We first perform the infinite sum by (generalized) $\zeta$-function and finally take the limit $d\to 0$, which turns out to be useful.

In the following, we will focus on the off-diagonal elements in the adjoint representation of $SU(N)$ as the diagonal ones do not couple to magnetic fluxes and therefore have different mass spectra. We will give some comments on the diagonal part separately in Sec.~\ref{diagonal elements}. The mass formulas for off diagonal components are summarized as follows:
\begin{align}
    m_{S,ab,(r_1,r_2,r_3)}^2=&\sum_{n=1}^3\frac{2\pi|M_n^{ab}|(2r_n+1)}{\mathcal A_n},\\
    m_{F,ab,(r_1,r_2,r_3),g}^2=&\mu_{g,ab}^2+\sum_{n=1}^3\frac{4\pi|M_n^{ab}|r_n}{\mathcal A_n},\\
    m_{V,ab,(r_1,r_2,r_3)}^2=&m_{S,ab,(r_1,r_2,r_3)}^2,\\
    m_{z^n,ab,(r_1,r_2,r_3)}^2=&\left(\sum_{m=1}^3\frac{2\pi|M_m^{ab}|}{\mathcal A_m}-\frac{4\pi M^{ab}_n}{\mathcal A_n}\right)+\sum_{m=1}^3\frac{4\pi|M_m^{ab}|r_m}{\mathcal A_m}\\
    m_{\bar{z}^n,ab,(r_1,r_2,r_3)}^2=&\left(\sum_{m=1}^3\frac{2\pi|M_m^{ab}|}{\mathcal A_m}+\frac{4\pi M^{ab}_n}{\mathcal A_n}\right)+\sum_{m=1}^3\frac{4\pi|M_m^{ab}|r_m}{\mathcal A_m},
\end{align}
of scalars, fermions, vectors $W_\mu^{ab}$, and $W_{z(\bar{z})}^{ab}$, respectively. The label $g$ of a fermion mass denotes its chirality on tori,\footnote{It is different from the gauge coupling $g$ in the original action, which would be understood from the context.} and $\mu_{g,ab}^2$ can be found e.g. from \eqref{kk+++}-\eqref{kk--+}.\footnote{Nevertheless, in the following discussion, we do not need to know the values of each $g$ as we mentioned before since we sum up all the contributions, which is independent of the signs of fluxes as well as the chirality assignments.} In general, fermions have zero modes and also $W_{z(\bar{z})}^{ab}$ may have zero modes depending on the background fluxes. Therefore, one has to carefully treat (possible) zero modes separately since naive integration of zero modes causes problematic behavior of the vacuum energy. 

We also summarize the number of degrees of freedom as follows: In the 10D case, the gauginos are described by Majorana-Weyl spinors, which can be decomposed as four 4D Weyl spinors and their conjugates.\footnote{We are able to decompose 10D $\mathcal{N}=1$ SYM multiplet into three chiral and one vector supermultiplets in 4D $\mathcal{N}=1$ supersymmetry.} The existence of zero modes depend on the sign of the flux difference $M^{ab}_n$ as we mentioned, whereas the massive modes always exist. For each mode (if exist), the degeneracy is given by $|M_1^{ab}M_2^{ab}M_3^{ab}|$. Nevertheless, in the following discussion, we will neglect the vacuum energy from zero modes since they do not depend on moduli fields.\footnote{This is not true in general when moduli are treated as dynamical fields. Since we are interested in the lowest order of the adiabatic expansion in moduli fields, namely constant moduli fields, the contribution to vacuum energy from zero modes will be discarded. Nevertheless, it is not difficult to include such contribution if necessary. } Therefore, we do not need to be careful about the difference of the sign of $M^{ab}_n$. For vector fields $W^{ab}_\mu$, each has three degrees of freedom as they are massive vectors. The origin of the longitudinal components is $W_{z^n(\bar{z}^n)}^{ab}$. Therefore, some part of the KK tower of $W_{z^n(\bar{z}^n)}^{ab}$ should be absorbed into $W^{ab}_\mu$ KK tower. Nevertheless, one does not need to take this fact into account by simply counting the degrees of freedom of $W^{ab}_\mu$ to be only two transverse modes. The contribution from the longitudinal modes can be taken account of by simply summing up the contribution from $W_{z^n(\bar{z}^n)}^{ab}$. Note also that all the bosonic components have $|M_1^{ab}M_2^{ab}M_3^{ab}|$ degeneracy. 

We regard the resultant vacuum energy as effective potential of moduli fields namely the area of tori. As we are interested in a generic point in moduli space, supersymmetry is broken except the points where a bosonic zero mode appears. The vacuum energy of the bosonic fields are summarized as
\begin{align}
    V_{B,ab}=-\frac{\Gamma\left(-\frac{d}{2}\right)\mu^{4}}{(4\pi)^{\frac d2}}\sum_{i=1}^8\zeta_B\left(-\frac d2,C_{B,i}^{ab}\biggr| C^{ab}_1,C^{ab}_2,C^{ab}_3\right),
\end{align}
  where we have defined
\begin{align}
    C_n^{ab}\equiv& \frac{4\pi|M_n^{ab}|}{\mu^2\mathcal A_n}.
\end{align}
Here, $\Gamma(z)$ is the $\Gamma$-function and $\zeta_B(s,w|a_1,a_2,a_3)$ is the Barnes $\zeta$-function~\eqref{def:barnesZ}. This expression is yet formal and we need to take the limit $d\to4$, which requires the analytic continuation of the Barnes $\zeta$-function. We have also introduced $\{C_{B,i}^{ab}\}$ as a set of the (dimensionless) lowest mass square in each KK tower, which can be summarized as
\begin{align}
C_s^{ab}\pm C_1^{ab},\quad C_s^{ab}\pm C_2^{ab},\quad C_s^{ab}\pm C_3^{ab}, C_s^{ab} \ (\times 2),
\end{align}
where $(\times 2)$ means that there are two KK towers with such a lowest mass square and
\begin{align}
     C_s^{ab}\equiv &\frac12\left(C_1^{ab}+C_2^{ab}+C_3^{ab}\right).\label{defCs}
\end{align}
Similarly, the fermionic vacuum energy is given by
\begin{align}
  V_{F,ab} = \frac{\Gamma\left(-\frac{d}{2}\right)\mu^{4}}{(4\pi)^{\frac d2}}\sum_{i=1}^8\zeta_B\left(-\frac d2,C_{F,i}^{ab}\biggr| C^{ab}_1,C^{ab}_2,C^{ab}_3\right),
\end{align}
where $\{C_{F,i}^{ab}\}=\{(\mu_g^{ab})^2/\mu^2\}$, which are explicitly given by
\begin{align}
    0,\quad C_1^{ab},\quad C_2^{ab},\quad C_3^{ab},\quad C_1^{ab}+C_2^{ab}, \quad C_2^{ab}+C_3^{ab}\quad C_3^{ab}+C_1^{ab}, \quad C_1^{ab}+C_2^{ab}+C_3^{ab}
\end{align}
 We note that the values of $\{C_{B,i}^{ab}\}$ and $\{C_{F,i}^{ab}\}$ are completely different from each other unless there appears a zero mode in the bosonic sector, and therefore, cancellation between bosons and fermions is not expected in general.

\subsection{Absence of UV divergences}
Let us discuss the UV part of the vacuum energy. The vacuum energy is formally given by
\begin{align}
    V_{\text{1-loop}}=-\frac{\Gamma\left(-2+\epsilon\right)\mu^{4}}{(4\pi)^{2-\epsilon}}\sum_{i=1}^8\biggl[\zeta_B\left(-2+\epsilon,w\biggr| C^{ab}_1,C^{ab}_2,C^{ab}_3\right)\biggr]^{C_{B,i}^{ab}}_{C_{F,i}^{ab}},
\end{align}
where we have introduced a notation $[f(w)]^{a}_b=f(a)-f(b)$.
We emphasize that the above formula is a formal one since the fermionic sector has a zero mode, or more precisely the KK tower starting from a zero mode. Strictly speaking, the Barnes $\zeta$-function $\zeta_B(s,w|a,b,c)$ is not well defined when $w=0$. Such a subtlety also appears for a bosonic KK tower at a supersymmetric point. Nevertheless, if we treat such modes as if they were well defined, we find an interesting UV property: One can find the UV divergent part proportional to $1/\epsilon$, which can be written as
\begin{align}
    V^{\rm UV}_{\text{1-loop}}=&\frac{\mu^{4}}{(4\pi)^{2}C_1^{ab}C_2^{ab}C_3^{ab}\epsilon}\sum_{k=0}^4\frac{\hat{B}_{3,k}}{(5-k)!k!}\sum_{i=1}^8\left[
    (C_{B,i}^{ab})^{5-k}- (C_{F,i}^{ab})^{5-k}\right],
\end{align}
where $\hat{B}_{3,k}$ is defined through
\begin{align}
    t^{3}\prod_{m=1}^3\frac{C_m^{ab}}{e^{C_m^{ab}t}-1}=\sum_{k=0}^{\infty}\hat{B}_{3,k}\frac{t^k}{k!},
\end{align}
and more explicitly
\begin{align}
    \hat{B}_{3,0}=&1,\\
    \hat{B}_{3,1}=&-\frac12\sum_{n=1}^3C_n^{ab},\\
    \frac{1}{2!}\hat{B}_{3,2}=&\frac{1}{12}((C^{ab}_1)^2+(C_2^{ab})^2+(C_3^{ab})^2+3C_1^{ab} C_2^{ab}+3C_2^{ab}C_3^{ab}+3C_1^{ab}C_3^{ab}),\\
   \frac{1}{3!} \hat{B}_{3,3}=&-\frac{1}{24}\left(C_1^{ab}C_2^{ab}(C_1^{ab}+C_2^{ab})+\text{cyclic}\right)-\frac18 C_1^{ab}C_2^{ab}C_3^{ab},\\
   \frac{1}{4!} \hat{B}_{3,4}=&\frac{1}{720}\left(\sum_{n=1}^3\left\{-(C_n^{ab})^4+5\frac{\prod_{m=1}^3(C_m^{ab})^2}{(C_n^{ab})^2}\right\}+15\left(\prod_{m=1}^3C_m^{ab}\right)\left(\sum_{n=1}^3C_n^{ab}\right)\right).
\end{align}
One can confirm that the direct substitution of $\{C_{B(F),i}^{ab}\}$ and the above coefficients read exactly
\begin{align}
    V^{\rm UV}_{\text{1-loop}}=0.
\end{align}
Thus, we find that the UV divergence of the vacuum energy exactly cancels if we allow to use the formal Barnes $\zeta$-function, which is ill-defined for a KK tower including a zero mode. Note that, as is clear from explicit formulas for $\{C_{B(F),i}^{ab}\}$, such a cancellation is highly non-trivial, and the UV divergence vanishes if and only if we sum up all the contribution. Nevertheless, the use of the ill-defined Barnes $\zeta$-function leads to problems for remaining terms. Therefore, we will not take this prescription. 

The only way to avoid the subtlety of the ill-defined Barnes $\zeta$-function is to decompose the KK tower as 
\begin{align}
    (r_1,r_2,r_3)=&(0,0,0),(\tilde{r}_1,0,0),(0,\tilde{r}_2,0),(0,0,\tilde{r}_3),\nonumber\\
    &(0,\tilde{r}_2,\tilde{r}_3),(\tilde{r}_1,0,\tilde{r}_3),(\tilde{r}_1,\tilde{r}_2,0),(\tilde{r}_1,\tilde{r}_2,\tilde{r}_3),\label{decompose}
\end{align}
where $1\leq\tilde{r}_i\in{\mathbb N}$. With such a decomposition, we are able to avoid to use the ill-defined Barnes $\zeta$-function. Naively, this decomposition is necessary only for the KK tower including a zero mode. However, in order to make supersymmetry manifest at supersymmetric points, one has to decompose all the modes in the same way. Then, the vacuum energy is expressed as
\begin{align}
    \tilde{V}_{\text{1-loop}}=&-\frac{\Gamma\left(-\frac d2\right)\mu^4}{(4\pi)^{\frac d2}}\sum_{i=1}^8\Biggl[w^{\frac d2}+\sum_{n=1}^{3}\biggl\{(C_n^{ab})^{\frac d2}\zeta_H\left(-\frac d2, \frac{w}{C_n^{ab}}+1\right)\nonumber\\
    &+\zeta_B\left(-\frac{d}{2},w+2C_s^{ab}-C_n^{ab}\biggr|C_{p\neq n}^{ab},C_{q\neq n}^{ab}\right)\biggr\}+\zeta_B\left(-\frac{d}{2},w+2C_s^{ab}\biggr|C_1^{ab},C_2^{ab},C^{ab}_3\right)\Biggr]_{C_{F,i}^{ab}}^{C_{B,i}^{ab}},
\end{align}
where $\zeta_H(s,w)$ is the Hurwitz $\zeta$-function, which can also be understood as a special case of the Barnes $\zeta$-function.  This expression manifests that the vacuum energy vanishes at supersymmetric points while avoiding the problem of the ill-defined Barnes $\zeta$-function.

Let us again discuss the UV divergent part of the effective potential. The $1/\epsilon$ pole in $d=4-2\epsilon$ with $\epsilon\to 0$ is 
\begin{align}
\tilde{V}^{\rm UV}_{\text{1-loop}}=&\frac{\mu^{4}}{(4\pi)^{2}\epsilon}\sum_{i=1}^8\Biggl[\sum_{k=0}^4\frac{\hat{B}_{3,k}}{C_1^{ab}C_2^{ab}C_3^{ab}(5-k)!k!}
   (w+2C_s^{ab})^{5-k}-\frac12 w^2\nonumber\\
   &\qquad +\sum_{n=1}^{3}\biggl\{\sum_{k=0}^2\frac{\hat{B}_{1,k}^{(n)}(w+C_n^{ab})^{3-k}}{C_n^{ab}(3-k)!k!}-\sum_{k=0}^3\frac{C_n^{ab}\hat{B}^{(n)}_{2,k}(w+2C_s^{ab}-C_n^{ab})^{4-k}}{C_1^{ab}C_2^{ab}C_3^{ab}(4-k)!k!}\biggr\}\Biggr]_{C_{F,i}^{ab}}^{C_{B,i}^{ab}},  
\end{align}
where $B^{(n)}_{2,k}$ is defined through
\begin{align}
    t^{2}\frac{e^{C_n^{ab}t}-1}{C_n^{ab}}\prod_{m=1}^3\frac{C_m^{ab}}{e^{C_m^{ab}t}-1}=\sum_{k=0}^{\infty}\hat{B}_{2,k}^{(n)}\frac{t^k}{k!}.
\end{align}
and $\hat{B}_{1,k}^{(n)}$ is defined through
\begin{align}
    \frac{C_n^{ab}t}{e^{C_n^{ab}t}-1}=\sum_{k=0}^{\infty}\hat{B}_{1,k}^{(n)}\frac{t^k}{k!}.
\end{align}

We have computed the divergent vacuum energy (with Mathematica), and again found 
\begin{align}
   \tilde{V}^{\text{UV}}_{\text{1-loop}}=0.
\end{align}
The bosonic contribution is given by
\begin{align}
    V_B=&\frac{\mu^4}{960\pi^2\epsilon}\left(\frac{C_2^3+C_3^3}{C_1}+\frac{C_3^3+C_1^3}{C_2}+\frac{C_1^3+C_2^3}{C_3}\right)-\frac{\mu^4}{192\pi^2\epsilon}(C_1C_2+C_2C_3+C_3C_1),\label{UVdivergence}
\end{align}
and the fermionic contribution has precisely the same form with an opposite sign, which cancels the bosonic contribution.\footnote{We comment on KK tower decomposition~\eqref{decompose} we have used. Since only the fermionic KK tower with a zero mode makes the Barnes $\zeta$-function ill-defined, one may think that the decomposition~\eqref{decompose} is necessary only for such a tower. We have examined to decompose the fermionic KK tower and one of the bosonic KK towers that has a possible zero mode, and computed the UV divergent terms. However, in such a case, we found no cancellation of the UV divergences. On the other hand, as we have seen, when we decompose all the KK towers in the same way, we have found an exact cancellation of the UV divergence. Therefore, we believe that our decomposition~\eqref{decompose} would be more suitable than the partial decomposition, which keeps underlying $\mathcal{N}=4$ supersymmetry mentioned above.}

We emphasize that the cancellation of the UV divergence is independent of the moduli configuration. Therefore, even if 4D $\mathcal{N}=1$ supersymmetry is spontaneously broken, there is no UV divergent vacuum energy. This is highly nontrivial as the bosonic and fermionic masses completely differ from each other and there is no mass degeneracy. Recall that the divergent vacuum energy in field theory is formally given by $V=a_0\Lambda^4{\rm STr}\ m^0+a_1\Lambda^2{\rm STr}\ m^2+a_2{\rm STr}\{\log(m^2/\Lambda^2)\ m^4\}$
where $a_{0,1,2}$ are numerical factors, $\Lambda$ being a divergent cut-off, and ${\rm STr}$ denotes the super-trace and $m$ denotes masses of fields. 4D ${\cal N}=1$ supersymmetry ensures ${\rm STr}\ m^0$ even if the symmetry is spontaneously broken, but in general ${\rm STr}\ m^2,{\rm STr}\ m^4$ are non-vanishing in 4D $\mathcal{N}=1$ supersymmetric theory if the symmetry is spontaneously broken. However, in our case, the log divergence is absent, which implies that ${\rm STr}\ m^4=0$.\footnote{The dimensional regularization excludes the first two contribution, and we cannot conclude the absence of the quadratic divergences, in general. However, 4D $\mathcal{N}=1$ supersymmetric case, even if spontaneously broken, the quadratic divergences are absent in general. Thus, only the logarithmic divergences matter, which can be dealt with the dimensional regularization.}
This is a consequence of  $\mathcal{N}=4$ supersymmetry {\it spontaneously} broken by the magnetic fluxes which can be confirmed by vanishing supertrace ${\rm STr} (m^{2n})=0$ for $n<4$~\cite{Bachas:1995ik}. However, as we will show below, the model with 4D $\mathcal{N}=1$ supersymmetric completions, which seems to break the structure of extended supersymmetry explicitly, yields the UV finite one-loop vacuum energy as this case. Such a behavior is possible if the KK mass spectrum of the model with and without the completions are the same, which implies that the 4D $\mathcal{N}=1$ supersymmetric completions change the tree level potential without affecting the KK masses.

\subsection{The behavior of the 1-loop effective potential}
The full formulas of the vacuum energy is rather complicated, which requires us to use the analytically continued Barnes $\zeta$-function. Explicitly, it is given by
\begin{align}
    \tilde{V}_{\text{1-loop}}=&\lim_{\epsilon\to0}-\frac{\mu^4}{(4\pi)^{2-\epsilon}}\sum_{i=1}^8\Biggl[\Gamma\left(-2+\epsilon\right)w^{2-\epsilon}+\sum_{n=1}^{3}\biggl\{ \sum_{k=0}^2\frac{(-1)^k\hat{B}^{(n)}_{1,k}\Gamma\left(-3+k+\epsilon\right)}{k!C_{n}^{ab}}(w+C_n^{ab})^{3-k-\epsilon}\nonumber\\
&+\sum_{k=0}^3\frac{(-1)^k\hat{B}^{(n)}_{2,k}C_n^{ab}\Gamma\left(-4+k+\epsilon\right)}{k!C_{1}^{ab}C_{2}^{ab}C_{3}^{ab}}(w+2C_s^{ab}-C_n^{ab})^{4-k-\epsilon}
\biggr\}\nonumber\\
&+\sum_{k=0}^4\frac{(-1)^k\hat{B}_{3,k}\Gamma\left(-5+k+\epsilon\right)}{k!C_1^{ab}C_{2}^{ab}C_{3}^{ab}}(w+2C_s^{ab})^{5-k-\epsilon}\Biggr]_{C_{F,i}^{ab}}^{C_{B,i}^{ab}}\nonumber\\
&-\frac{\mu^4}{(4\pi)^{2}}\int_0^\infty dt\sum_{i=1}^8\Biggl[\sum_{n=1}^3\Biggl\{e^{-wt}\left(\frac{t^{-3}}{(1-e^{-C_n^{ab}t})}-\frac{t^{-4}}{C_n^{ab}}\sum_{k=0}^2\frac{(-1)^k\hat{B}_{1,k}^{(n)}t^k}{k!}\right)\nonumber\\
&+ e^{-(w+2C_s^{ab}-C_n^{ab})t}\left(\frac{(1-e^{-C_n^{ab}t})t^{-3}}{\prod_{l=1}^3(1-e^{-C_l^{ab}t})}-\frac{C_n^{ab}t^{-5}}{\prod_{l=1}^3C_l^{ab}}\sum_{k=0}^3\frac{(-1)^k\hat{B}_{2,k}^{(n)}t^k}{k!}\right)\Biggr\}\nonumber\\
&+e^{-(w+2C_s^{ab})t}\left(\frac{t^{-3}}{\prod_{l=1}^3(1-e^{-C_l^{ab}t})}-\frac{t^{-6}}{\prod_{l=1}^3C_l^{ab}}\sum_{k=0}^4\frac{(-1)^k\hat{B}_{3,k}t^k}{k!}\right)\Biggr]_{C_{F,i}^{ab}}^{C_{B,i}^{ab}}.\label{fullpotential}
\end{align}
The first three lines are regular in the limit $\epsilon\to0$ as we have discussed, and can be simplified further. The remaining terms contain an integral over an auxiliary parameter $t$, which is necessary to render the asymptotic series well-defined. 

We have a few technical comments: Although the effective potential is finite, the terms including $t$-integration makes it difficult to have a handy analytic formula. We would like to comment on the difficulty to get an analytical result in some limit: One may wonder if the large volume limit $\mathcal{A}_n\to\infty$ simplifies the formula. However, one immediately notice that the integrands can be written by the ratio of the flux dependent masses and simultaneous large volume limit for all tori do not lead to any simplification. Furthermore, if one considers an asymmetric limit e.g. taking one of the tori to be large or small. However, such a limit is problematic since there appear  tachyonic modes in the bosonic sector. Then, the effective potential acquires an imaginary part corresponding to the instability due to the tachyonic mode. Note also that the effective potential is independent of the renormalization scale, which can be confirmed by changing the integration variable $t$. Thus, we will discuss the behavior of the effective potential numerically in the following. We give related comments on the dimensional reduction in magnetized tori models in Appendix~\ref{dimred}.

The Barnes $\zeta$-function method is applied to the case of the superspace completion models in the exactly the same way, but as we will show in an explicit example, the KK towers differ from the ones considered this section.
\subsection{Notes on diagonal elements}\label{diagonal elements}
The diagonal elements in the adjoint representation of $U(N)$ do not couple to any background fluxes. Therefore, their KK mass spectra are given by the eigenvalues of the Laplacian on $({\mathbb T}^2)^n$ independently of the background fluxes we introduce.\footnote{Strictly speaking, this is not true as the flux and the vacuum energy may deform background geometry, which modify the KK mass spectra of fields that do not directly couple to the background fluxes. This may not be the case if the dimensionful gauge coupling $g$ is small enough. We here simply assume that the background geometry is well approximated by the original tori for any flux backgrounds.} In general, the KK mass for tori without flux is given by (see e.g.~\cite{Hamada:2012wj}),
\begin{align}
    m_{r_{R,n},r_{I,n}}^2=\frac{4\pi^2{\rm Im}\tau_n}{\mathcal{A}_n}\left(r_{R,n}^2+\left(\frac{1}{{\rm Im}\tau_n}\right)^2(r_{I,n}-r_{R,n}{\rm Re}\tau_n)^2\right),\label{nofluxmass}
\end{align}
where $r_{R,I}\in {\mathbb Z}$.
In particular, when the tori are flat, ${\rm Im}\tau_i=1$, the formula takes a simple form $m^2= \sum_n\frac{1}{R_n^2}(r^2_{R,n}+r_{I,n}^2)$. This formula applies to the off-diagonal elements if the effective flux $M^{ab}_n$ is zero.

As the number and the KK mass spectra of bosonic and fermionic states of diagonal elements are precisely the same, they cancel each other, which is a consequence of residual supersymmetry independently of the flux configuration. Therefore, such contribution can be negligible within our setup.\footnote{This is not true if e.g. gaugino masses are lifted due to supersymmetry breaking in some sector. Then, complete cancellation fails and there would be a vacuum energy proportional to the soft gaugino mass. Such a situation is generally expected to occur due to anomaly mediation. }

\section{An illustrating example}\label{example}
\subsection{A model without 4D $\mathcal{N}=1$ supersymmetric completions}\label{exwo}
\begin{figure}[htbp]
    \centering
\includegraphics[keepaspectratio, scale=0.8]{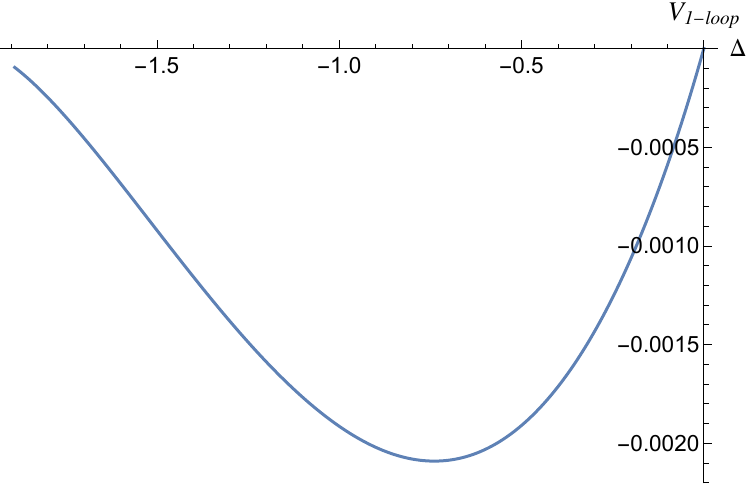}
    \caption{The effective potential in the $U(2)$ model. We have taken the parametrization~\eqref{ex-para} with parameters $f=1$, and taken the renormalization scale $\mu=1$ as a unit.}
    \label{fig:Vwide}
    \end{figure}

\begin{figure}[htbp]
    \centering
\includegraphics[keepaspectratio, scale=0.8]{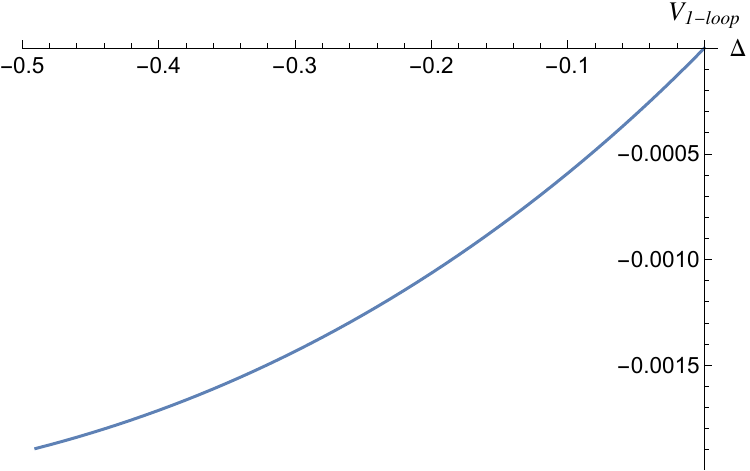}
    \caption{The effective potential in the $U(2)$ model near $\Delta=0$. The parameters are the same as that in Fig~\ref{fig:Vwide}.}
    \label{fig:Vsmall}
    \end{figure}

We discuss the behavior of the vacuum energy within concrete models of 10D SYM. we consider $U(2)$ gauge theory with the background flux:
\begin{equation}
   {\bm M}_{n}=\left(\begin{array}{cc}f&0\\ 0&0\end{array}\right),\label{U2breaking}
\end{equation}
for all $n=1,2,3$ where $f>0$. In this case, $M^{12}_n=f=-M^{21}_n$.  In this case, the lowest mass eigenstate is one of $\phi^{12}_{z^n,(0,0,0)}$ ($n=1,2,3$), which has
\begin{align}
    m_{z^n,12,(0,0,0)}^{2}=\sum_{m=1}^3\frac{2\pi f}{\mathcal{A}_m}-\frac{4\pi f}{\mathcal{A}_n},
\end{align}
and therefore, there appears a massless scalar for $\mathcal{A}_1:\mathcal{A}_2:\mathcal{A}_3=1:2:2$ or for its cyclic permutations, which are supersymmetric hypersurfaces in the moduli space. In this case, the number of zero modes is given by $f$, and therefore, we simply multiply the factor to the potential~\eqref{fullpotential}, and only the $(12)$ sector feels the magnetic fluxes. We are interested in the supersymmetric vacua, and we parametrize
\begin{align}
    C_1^{12}=2+\Delta,\quad C_2^{12}=C_3^{12}=1,\label{ex-para}
\end{align}
where $\Delta\leq0$ corresponds to a modulus. As an illustration, we show the shape of the effective potential in Figs.~\ref{fig:Vwide}, \ref{fig:Vsmall}. Note that since $f$ just changes the overall scaling of the effective potential, and does not affect the shape of the potential. From the figures, we find that the supersymmetric point $\Delta=0$ is stable. Note also that $\Delta>0$ leads to a tachyonic mode, and effective potential would acquire an imaginary part. Note that we have fixed other moduli $C_2^{12},C_3^{12}$ corresponding to the areas $\mathcal{A}_{2,3}$, but if we turn on them as dynamical variables, the effective potential would give us supersymmetric surfaces in the moduli space, and the seemingly singular point $\Delta=0$ would become a point on the surfaces.


We would like to comment on some technical details in drawing the figures. We have taken the limit $\epsilon\to 0$ in \eqref{fullpotential}, which yields finite terms, and taken account of them exactly. For the integral terms, we have divided the integration region from $[0,0.2]$ and $[0.2,\infty)$. For the former region, we have expanded the integrand in powers of $t$ up to the $t^{20}$ order, and performed numerical integration at various $\Delta$. For the latter region, we have used the exact one in the numerical integration, which has a better convergence property thanks to the exponential decay.\footnote{We have integrated $t\in [0.2,20]$ in the numerical integration, but change of the upper limit does not change the result.} We have computed discrete 190 points and joined them. All the computation is done with Mathematica. We also note that we have checked that the results do not depend on the integer $M$, which appears in the regularization of the Barnes $\zeta$-function as it should be.

As clearly seen in Figs~\ref{fig:Vwide},~\ref{fig:Vsmall} the one-loop vacuum energy prefers to the supersymmetry breaking $\Delta\neq 0$. It is rather surprising that the one-loop correction leads to radiative supersymmetry breaking at first glance. However, this is not yet conclusive as we have not added the tree level contribution. Then, we also notice some important facts: Although we have naively reduced the ten-dimensional super-Yang-Mills theory with Abelian magnetic fluxes and confirmed possible supersymmteric mass spectra, it is not yet enough to realize four-dimensional $\mathcal{N}=1$ supersymmetric theory as we discuss in the next section.

\subsection{A model with 4D $\mathcal{N}=1$ supersymmetric completions}
We here discuss a model with 4D $\mathcal{N}=1$ supersymmetric completions. For 4D $\mathcal{N}=1$ supersymmetric completion models, existence of supersymmetric vacua depends on the signs of fluxes, and D-terms cannot be canceled among fluxes from each torus with the flux choice \eqref{U2breaking}. Therefore, we consider another flux parameters
\begin{align}
   {\bm M}_{1}=\left(\begin{array}{cc}\frac{f}{2}&0\\ 0&-\frac{f}{2}\end{array}\right),\quad {\bm M}_{2}={\bm M}_{3}=\left(\begin{array}{cc}-\frac{f}{2}&0\\ 0&\frac{f}{2}\end{array}\right),\label{SSfluxes}
\end{align}
which leads to the same numbers of zero modes $f$ $(f>0)$ since $|M^{12}_i|=f$ $(\forall i=1,2,3)$. We note that, even if we used \eqref{SSfluxes} instead of \eqref{U2breaking} to the model without 4D $\mathcal{N}=1$ supersymmetric completions, the result in the previous section does not change since the signs of fluxes do not really change the mass spectrum in the models without 4D $\mathcal{N}=1$ supersymmetric completions. With the above flux parameters, we have performed KK expansion of superfields in Appendix~\ref{KKexpansionSS}. We then find a set of KK masses for generic moduli space points as follows:\vspace{10pt}\\
\fbox{Fermion masses}
\begin{align}
   M^2_f=\left\{\begin{array}{ll}
   (0,0,0)& (\text{for }\phi_{1,\bm0}^+)\\[5pt]
  (r_1+1,r_2,r_3)&(\text{for  }\Phi^+_{1,\bm r}, \lambda_{\bm r}^+)\\[5pt]
  (r_1,r_2+1,r_3+1)& (\text{for }\Phi^-_{1,\bm r}, \lambda_{\bm r}^-)\\[3pt]
   (r_1,r_2,r_3)& (\text{for }\Phi^-_{2,\bm r}, \Phi^+_{3,\bm r})\\[5pt]
   (r_1+1,r_2+1,r_3+1)&(\text{for }\Phi^+_{2,\bm r}, \Phi^-_{3,\bm r})
   \end{array}
   \right.\label{SSMF2}
\end{align}
\fbox{Boson masses}
\begin{align}
    M^2_b=\left\{\begin{array}{ll}\left(-\frac12,\frac12,\frac12\right) &(\text{for }\phi_{1,\bm 0}^+)\\[5pt]
    \left(r_1+\frac12,r_2+\frac12,r_3+\frac12\right)&(\text{for }V_{\bm r}^+, \Phi^\pm_{1,\bm r})\\[5pt]
    \left(r_1-\frac12,r_2+\frac12,r_3+\frac12\right)&(\text{for }\Phi_{3,\bm r}^+)\\[5pt]
   \left(r_1+\frac12,r_2-\frac12,r_3-\frac12\right)&(\text{for }\Phi_{2,\bm r}^-)\\
\left(r_1+\frac12,r_2+\frac32,r_3+\frac32\right)&(\text{for }\Phi_{2,\bm r}^+)\\[5pt]
    \left(r_1+\frac32,r_2+\frac12+,r_3+\frac12\right)&(\text{for }\Phi_{3,\bm r}^-)\\[5pt]
    \end{array}\right.\label{SSMB2}
\end{align}
We have introduced a vector notation $(r_1,r_2,r_3)$ corresponding to $M_{\bm r}^2=\sum_{i=1}^3\frac{4\pi f r_i}{\mathcal{A}_i}$.
Note that from \eqref{Phi23cond}, the fermionic zero mode is only that in $\phi_{1,\bm 0}^+$. Supersymmetry is restored e.g. when $\mathcal{A}_2=\mathcal{A}_3=2\mathcal{A}_1$, which corresponds to the vanishing D-term condition
\begin{align}
    \frac{4\pi f}{g_{\rm 4D}\mathcal{A}_1}=\sum_{i=2}^3\frac{4\pi f}{g_{\rm 4D}\mathcal{A}_i},\label{SUSYcondition}
\end{align}
which minimizes the flux D-term potential~\eqref{fluxpotential}. Notice that when \eqref{SUSYcondition} is satisfied, flux induced D-term vanishes and the mass terms originate only from superpotential, which manifests supersymmetric KK mass spectrum. Nevertheless, the KK towers look different from that without 4D $\mathcal{N}=1$ completion, which should result in the difference of the one-loop vacuum energy. Practically, in computing the one-loop vacuum energy, one has to be careful about the condition~\eqref{Phi23cond}, which makes ``irregular'' the KK towers of $\Phi^+_{3,\bm r}$ and $\Phi_{2,\bm r}^-$ and we show them explicitly in the appendix~\ref{irregKK}. The vacuum energy contributions from other multiplets are computed in the same way as done in the previous sections.

\begin{figure}[htbp]
    \centering
\includegraphics[keepaspectratio, scale=0.9]{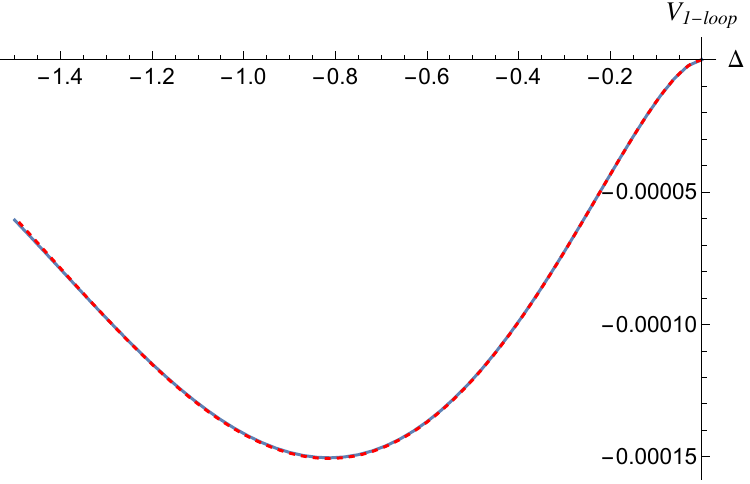}
    \caption{The comparison of the one-loop effective potential in the $U(2)$ model with (blue solid) and without (red dashed) 4D $\mathcal{N}=1$ supersymmetric completions. We have taken $f=1$ and $\mu=1$ as a unit. As we see, the two lines degenerate, which indicates that the KK mass spectra of two models are the same.}
    \label{fig:comparison}
    \end{figure}

We first focus on the UV divergent part of the one-loop effective potential. We perform the regularization with Barnes $\zeta$-function as is done previously. Rather surprisingly, the one-loop effective potential in the model with the 4D $\mathcal{N}=1$ supersymmetry completions is UV finite as the case without the completions. Furthermore, we have numerically computed the finite part of the one-loop effective potential and compared it with that in the model without the completions in Fig.~\ref{fig:comparison}, which shows that the two approach gives precisely the same effective potential. This result indicates that despite seemingly different KK tower structures the two models have the same KK particles by shuffling the KK towers in a nontrivial way. Unfortunately, we have not succeeded in proving the equivalence of the KK towers, but the above result strongly supports the above argument. We however emphasize that it is yet true that the two models are physically different since the tree level flux induced potential in the two models are different. This indicates that the 4D $\mathcal{N}=1$ supersymmetry completions only changes some part of the D-term potential without affecting KK mass spectra. As far as we know, the differences between the component action and the superspace-embedded one have not been fully addressed in the presence of nontrivial background fields such as magnetic fluxes. It would be worth investigating it further, which we will leave for future work.

We finally give a comment to the relation to the previously known result~\cite{Buchmuller:2019zhz,Buchmuller:2020nnl}. It is worth noting that the authors of~\cite{Buchmuller:2019zhz,Buchmuller:2020nnl} found that their field theory calculation based on superspace action shows the correspondence to the field theory limit of one-loop string calculation. Such a result implies that the string theory implicitly knows the superspace completion automatically, which would result in the correspondence they found. It would be an interesting question, if realizable in string theory, whether the string version of $U(N)$ model considered in this paper gives the same result as that in the superspace completion model when a field theory limit is taken. This issue is beyond the scope of this work and we leave it for future work.  

\section{Conclusion}\label{conclusion}
In this work, we have studied the one-loop vacuum energy in 10D super Yang-Mills theory compactified on tori with Abelian magnetic flux background. We have considered two models that differ by 4D $\mathcal{N}=1$ supersymmetric completions. Since the KK mass structure is of the form of the Landau level, we have applied regularization of infinite KK mode sum by using Barnes $\zeta$-function regularization. Such a technique is also applicable to compute e.g. the corrections to gauge couplings from KK modes~\cite{Shibasaki:2021,Nakano:2023} as well. In the case without 4D $\mathcal{N}=1$ supersymmetric completions, we have obtained UV finite one-loop vacuum energy (independently of whether KK mass spectrum is supersymmetric or not), which can be understood as the consequence of spontaneously broken extended supersymmetry. We note however that, despite a good UV behavior of radiative corrections, the tree level flux induced potential is positive definite. Therefore, it is unlikely to have realistic vacua with almost vanishing vacuum energy in the model without 4D $\mathcal{N}=1$ supersymmetric completions. 

 We have also discussed the model with 4D $\mathcal{N}=1$ supersymmetric completions, in which we first embed 10D super Yang-Mills theory into 4D $\mathcal{N}=1$ superfields and introduce background magnetic fluxes. In this model, the flux induced D-term potential may vanish with appropriate choices of fluxes and areas of tori, and we expect more realistic vacua having (almost) vanishing vacuum energy. In the same way as the model without completions, we have performed infinite KK sum and found the same UV finite one-loop effective potential as in the case without the completions, which indicates that the seemingly different KK mass spectra in the two models are the same although its proof is rather nontrivial. This result seems rather surprising since the two models are yet different from each other by the flux induced tree level potential.

Our treatment of the vacuum energy as functions of volume moduli is yet unsatisfactory since we have not fully included all supergravity corrections. Nevertheless, we expect that possible supergravity corrections do not change our result in certain limits such as decoupling limit of Planck scale. An extension of our analysis to supergravity would be an interesting future direction of study.


We have possible extensions of our work other than that mentioned above. A possible straightforward extension is the magnetized orbifold models~\cite{Braun:2006se,Abe:2008fi,Fujimoto:2013xha,Abe:2013bca} in which KK spectrum differs from ours. In particular, such singular conditions in general reduce the number of supercharges, and therefore, the vacuum energy would behave differently. Another possible generalization is effective theory of magnetized $Dp$-brane models with $p\leq9$, where the field theory on the brane spans only in the lower dimensions. Combination of several branes~\cite{Abe:2015jqa} would be useful to realize standard model like spectrum without exotic charged matters, and similar consideration as ours would be important to discuss the moduli stabilization in such models~\cite{Abe:2017gye}. It would also be important to consider connections to magnetic compactifications of string theory, which has been addressed in SO(32) model in~\cite{Buchmuller:2019zhz,Buchmuller:2020nnl}. In our work, we have treated just a field theory rather than string theory, and stringy consistency condition further imposes some restriction to the model building. We expect that some of our results would be reproduced as the field theory limit of string theory, e.g.~\cite{DiVecchia:2011mf}, but nevertheless it would be important to consider stringy effects. We will leave these issues for future work.

\section*{Acknowledgement}
The authors would like to thank H.~Nakano for discussions about the usefulness of Barnes $\zeta$-function in the framework of multiple magnetized tori. 
YY is supported by Waseda University Grant for Special Research Projects (Project number: 2023C-584).
\appendix

\section{Barnes $\zeta$-function}\label{Barneszeta}
The Barnes $\zeta$-function is defined as
\begin{align}
    \zeta_B(s,w|a_1,\cdots,a_N)\equiv \sum_{n_1,\cdots,n_N\geq 0}\frac{1}{(w+a_1n_1+\cdots+a_Nn_N)^s}\label{def:barnesZ}
\end{align}
for ${\rm Re}\ a_i>0, \ {\rm Re}\ s>N$. Analytic continuation of the Barnes $\zeta$-function to smaller value of ${\rm Re}\ s$ is possible and is given by
\begin{align}
  \zeta_B(s,w|a_1,\cdots,a_N)=&\frac{1}{\Gamma(s)\prod_{l=1}^Na_l}\sum_{k=0}^M\frac{(-1)^kB_{N,k}(0)\Gamma(s-N+k)}{k!w^{s-N+k}}\nonumber\\
  &+\frac{1}{\Gamma(s)}\int^{\infty}_0dt e^{-wt}\left(\frac{t^{s-1}}{\prod^N_{l=1}(1-e^{-a_l t})}-\frac{t^{s-N-1}}{\prod_{l=1}^N a_l}\sum^{M}_{k=0}(-)^kB_{N,k}(0)\frac{t^{k}}{k!}\right)
\label{continuedBZ}
\end{align}
which is valid for ${\rm Re}\ w>0$, ${\rm Re}\ s>N-M-1$ assuming $M>N$. Here, the coefficients $B_{N,k}(0)$ is defined through the generating function
\begin{align}
   t^N e^{xt} \prod^N_{j=1}\frac{a_j}{(e^{a_jt}-1)}=\sum_{n=0}^\infty\frac{t^n}{n!}B_{N,n}(x).
\end{align}
Notice that the term in the second line of \eqref{continuedBZ} can be understood as the Laplace transformation of the following asymptotic series
\begin{align}
    f_{N,M}(a_1,a_2\cdots,a_N;w,s)=\frac{t^{s-N-1}}{\prod_{l=1}^N a_l}\sum_{k=M+1}^{\infty}B_{N,k}(0)\frac{(-t)^k}{k!}.
\end{align}
Therefore, performing $t$-integration in the second line of \eqref{continuedBZ} yields a formal series
\begin{align}
  I=&\frac{1}{\prod_{l=1}^N a_l}\sum_{k=M+1}^{\infty}B_{N,k}(0)\frac{\Gamma(s-N+k)w^{-s+N-k}}{k!}\nonumber\\
  =&\frac{1}{\prod_{l=1}^N a_l}\sum_{k=1}^{\infty}B_{N,k+M}(0)\frac{\Gamma(M-N+s+k)}{(k+M)!w^{M-N+s+k}}.
\end{align}
Note that ${\rm Re}(M-N+s+k)>0$ for $k\geq1$ by assumption. The formal series is in general an asymptotic series in $1/w$, and only the Laplace transform expression makes sense. 
Note also that the Hurwitz $\zeta$-function defined as
\begin{align}
    \zeta_H(s,a)=\sum_{n=0}^\infty\frac{1}{(n+a)^s}
\end{align}
where $0<a<1$, ${\rm Re}\ s>1$, is regarded as a special case of the Barnes $\zeta$-function with $N=1$.

\section{KK expansion on superspace}\label{KKexpansionSS}
In this section, we discuss KK expansion of superfields shown in Sec.~\ref{SSdescription}. The procedure of the KK expansion is almost the same as that for components discussed in Sec.~\ref{withoutSS}. In the following, we assume $M^{ab}_i>0$.
Defining $a_{abi}=\sqrt{\frac{\mathcal{A}_i}{4\pi|M_i^{ab}|(\pi R_i)^2}}\bar{D}^{ab}_{\bar i}$ and $a^\dagger_{abi}=\sqrt{\frac{\mathcal{A}_i}{4\pi|M_i^{ab}|(\pi R_i)^2}}D^{ab}_i$, they satisfy
$[a_{i}^{(ab)},a^{(ab)\dagger}_{j}]=\delta_{{\bar i}j}$. Note that for $M^{ab}_i<0$ we need to change the role of $a$ and $a^\dagger$ and we identify $a_{abi}=-\sqrt{\frac{\mathcal{A}_i}{4\pi|M_i^{ab}|(\pi R_i)^2}}D^{ab}_i$ and $a_{abi}^\dagger=-\sqrt{\frac{\mathcal{A}_i}{4\pi|M_i^{ab}|(\pi R_i)^2}}\bar{D}^{ab}_{\bar i}$.\footnote{Essentially, this definition exchanges the role of $\bar{z}^i$ and $z^i$ of $(ba)$-sector having $M^{ba}>0$ and therefore, corresponding mode functions become complex conjugate to each other. More explicitly, for $M^{ab}_i>0$ a KK tower is constructed with a function satisfying $(\bar{\partial}_{\bar i}+\frac{\pi}{2{\rm Im}\tau_i}|M^{ab}_i|z^i)\chi^{0,(i)}_{(ab)}=0$ whereas $ba$-sector, we use $(\partial_i+\frac{\pi}{2{\rm Im}\tau_i}|M^{ab}_i|\bar{z}^i)\chi^{0,(i)}_{(ba)}=0$ and therefore, we find a relation $\chi^{0,(i)}_{(ba)}=(\chi^{0,(i)}_{(ab)})^*$. } On each torus $T^i$, there are $|M_i^{ab}|$ zero mode functions satisfying $a_{i}^{(ab)}\chi_{(ab)g}^{0,(i)}(z^i,\bar{z}^i)=0$ ($g=1,2,\cdots,|M^i_{ab}|$) and $\int_{T_i} d^2z^i \sqrt{g_i}\bar{\chi}_{(ab)g}^{0,(i)}\chi_{(ab)h}^{0,(i)}=\delta_{gh}$ from which we obtain an orthonormal eigenfunctions $\{\chi_{(ab)g}^{r_i,(i)}(z^i,\bar{z}^i)\}$ defined by $\chi_{(ab)g}^{r_i,(i)}(z^i,\bar{z}^i)=\frac{1}{\sqrt n!}(a^{(ab)\dagger}_{i})^{r_i}\chi_{(ab)g}^{0,(i)}$ and satisfying $\int_{T_i} d^2z^i \sqrt{g_i}\bar{\chi}_{(ab)g}^{r_i,(i)}\chi_{(ab)h}^{p_i,(i)}=\delta_{gh}\delta_{r_i,p_i}$. Note that $\chi_{(ab)g}^{0,(i)}$ can be normalized solution if and only if $M^{ab}_i\geq0$. Note that for superfields that do not interact with the background magnetic flux $M^{ab}_i=0$, the wave function on the tori become simple sinusoidal functions. We do not show it explicitly here since we are interested only in the supermultiplets coupled to the magnetic fields but show its zero mode wave function $\eta^{0}(z,\bar{z})=\prod_{i=1}^3(\mathcal{A}_i)^{-\frac12}$ where $\mathcal{A}_i=\int_{T_i} d^2z^i\sqrt{g_i}$, which is necessary in performing integration of compact spaces of the tadpole of $V^{aa}$. 

In the following, we consider a specific choice of fluxes in within a $U(2)$ model
\begin{align}
   {\bm M}_{1}=\left(\begin{array}{cc}\frac{f}{2}&0\\ 0&-\frac{f}{2}\end{array}\right),\quad {\bm M}_{2}={\bm M}_{3}=\left(\begin{array}{cc}-\frac{f}{2}&0\\ 0&\frac{f}{2}\end{array}\right),
\end{align}
where $f>0$. We have chosen the fluxes such that $|M_n^{12}|=f$, which would yield the same KK mass spectrum as the model in Sec.~\ref{example}. A flux-induced D-term vanishes when $\mathcal{A}_2=\mathcal{A}_3=2\mathcal{A}_1$ as we will see more explicitly. The flux differences are $M_1^{12}=f=-M_2^{12}=-M_{3}^{12}$. We expand off-diagonal superfields as
\begin{align}
    V^{+}=&\sum_{\bm r}V_{\bm r}^{+}(x,\theta,\bar{\theta})\chi^{r_1,(1)}(z^1,\bar{z}^1)\bar{\chi}^{r_2,(2)}(z^2,\bar{z}^2)\bar{\chi}^{r_3,(3)}(z^3,\bar{z}^3),\\
    \phi_i^{+}=&\sum_{\bm r}\phi_{i,\bm r}^{+}(x,\theta,\bar{\theta})\chi^{r_1,(1)}(z^1,\bar{z}^1)\bar{\chi}^{r_2,(2)}(z^2,\bar{z}^2)\bar{\chi}^{r_3,(3)}(z^3,\bar{z}^3),\\
    \phi_i^{-}=&\sum_{\bm r}\phi_{i,\bm r}^{-}(x,\theta,\bar{\theta})\bar{\chi}^{r_1,(1)}(z^1,\bar{z}^1)\chi^{r_2,(2)}(z^2,\bar{z}^2)\chi^{r_3,(3)}(z^3,\bar{z}^3),
\end{align}
where we have introduced the notation $V^{+}\equiv V^{12}, \ \phi^{+(-)}_{i}\equiv\phi^{12(21)}_i$.
The K\"ahler and super-potential of charged fields integrated over compact spaces are given by (up to quadratic terms)
\begin{align}
K_{\rm 4D}=&\frac{1}{g^2}\sum_{\bm r}\Biggl[\sum_{i=1}^3\left(|{\phi}^{-}_{i,\bm r}|^2+|{\phi}^{+}_{i,\bm r}|^2\right)+\sum_{i=1}^3\frac{8\pi M^a_i\sqrt{v_6}}{\mathcal{A}_i}V_{\bm 0}^{a}+2M^2_r|V^+_{\bm r}|^2-\frac{1}{\sqrt{v_6}}(|\phi_{i,\bm r}^+|^2-|\phi_{i,\bm r}^-|^2)({V}^{1}_{\bm 0}-{V}^{2}_{\bm 0})\nonumber\\
&+\sqrt{2}\left\{\left((m_{1,r_1}\ \overline{\phi^+_{1,\bm r-1_1}}-m_{1,r_1+1}\phi^-_{1,\bm r+1_1})-\sum_{i=2}^3(m_{i,r_i+1}\ \overline{\phi^+_{i,\bm r +1_i}}-m_{i,r_i}\phi^-_{i,\bm r-1_i})\right)\overline{V^+_{{\bm r}}}+{\rm h.c.}\right\}\Biggr],\\
    W_{\rm 4D}=&\frac{1}{g^2}\sum_{\bm r}\Biggl[\left(m_{2,r_2+1}\phi^+_{1,\bm r+1_2}\phi_{3,\bm r}^--m_{2,r_2}\phi^-_{1,\bm r-1_2}\phi^+_{3,\bm r}\right)+\left(m_{3,r_3+1}\phi_{2,\bm r+1_3}^+\phi_{1,\bm r}^--m_{3,r_3}\phi_{2,\bm r-1_3}^-\phi_{1,\bm r}^+\right)\nonumber\\
&+\left(m_{1,r_1+1}\phi^-_{3,\bm r +1_3}\phi_{2,\bm r}^+-m_{1,r_1}\phi^+_{3,\bm r-1_1}\phi^-_{2,\bm r}\right)\Biggr],
\end{align}
where $\sqrt{v_6}\equiv\sqrt{\prod_{i=1}^3\mathcal{A}_i}$ associated with the zero mode wave function, we have rescaled $\phi_i^\pm\to (\pi R_i)\phi_i^\pm$, 
and $M_r^2=\sum_{i=1}^3m_{i,r_i+1/2}^2$ and $\bm r\pm 1_i$ denotes that the $i$-th KK number is increased/decreased by 1. Note that one has to take $\phi^\pm_{i,\bm r}\equiv0$ when any KK number becomes negative.\footnote{Under this rule, one may freely shift the KK sum, e.g. by negative simultaneous shift $\bm r\to\bm r -1_i$. }

In order for mass terms to be diagonalized, one can use new bases~\cite{Shibasaki:2021,Nakano:2023} 
\begin{align}
    \Phi^+_{1,\bm r}=&\frac{1}{M_{1,\bm r}^+}\left(\sum_{i=2}^3m_{i,r_i}\phi^+_{i,\bm r-1_i}-m_{1,r_1+1}\phi^+_{1,\bm r+1_1}\right),\\
    \Phi^-_{1,\bm r}=&\frac{1}{M_{1,\bm r}^-}\left(-\sum_{i=2}^3m_{i,r_i+1}\phi^-_{i,\bm r+1_i}+m_{1,r_1}\phi^-_{1,\bm r-1_1}\right),\\
    \Phi^+_{2,\bm r}=&\frac{1}{M_{2,\bm r}^+}\left(m_{3,r_3+1}\phi^+_{2,\bm r+1_3}-m_{2,r_2+1}\phi^+_{3,\bm r+1_2}\right),\\
    \Phi^-_{2,\bm r}=&\frac{1}{M_{2,\bm r}^-}\left(m_{2,r_2}\phi^-_{3,\bm r-1_2}-m_{3,r_3}\phi^-_{2,\bm r-1_3}\right),\\
    \Phi^+_{3,\bm r}=&\frac{M_{2,\bm r}^-}{M_{3,\bm r}^+}\left(\phi_{1,\bm r}^++\sum_{i=2}^3\frac{m_{i,r_i}m_{1,r_1}}{(M_{2,\bm r}^-)^2}\phi^+_{i,\bm r-1_1-1_i}\right),\\
    \Phi^-_{3,\bm r}=&\frac{M_{2,\bm r}^+}{M_{3,\bm r}^-}\left(\phi_{1,\bm r}^-+\sum_{i=2}^3\frac{m_{i,r_i+1}m_{1,r_1+1}}{(M_{2,\bm r}^+)^2}\phi^-_{i,\bm r+1_1+1_i}\right),
\end{align}
where we have defined $M_{1,\bm r}^+\equiv\sqrt{\sum_{i=2}^3m_{i,r_i}^2+m_{1,r_1+1}^2}$, $M_{1,\bm r}^-\equiv\sqrt{\sum_{i=2}^3m_{i,r_i+1}^2+m_{1,r_1}^2}$, $M_{2,\bm r}^+\equiv\sqrt{\sum_{i=2}^3m_{i,r_i+1}^2}$, $M_{2,\bm r}^-\equiv\sqrt{\sum_{i=2}^3m_{i,r_i}^2}$, $M_{3,\bm r}^+\equiv\sqrt{\sum_{i=1}^3m_{i,r_i}^2}$, $M_{3,\bm r}^-\equiv\sqrt{\sum_{i=1}^3m_{i,r_i+1}^2}$. Notice that $\phi_{1,\bm 0}^+$ is not contained any of them and also that 
\begin{align}
    \Phi^-_{2,\bm r}=\Phi^+_{3,\bm r}\equiv 0 \quad (\text{for }r_2=r_3=0).\label{Phi23cond}
\end{align}
With the new basis, the 4D effective K\"ahler and super-potential can be rewritten as
\begin{align}
    K_{\rm 4D}=&\sum_{\bm r}\left(\sum_{i=1}^3(|\Phi^+_{i,\bm r}|^2+|\Phi^-_{i,\bm r}|^2)+2M^2_r|V^+_{\bm r}|^2-\sqrt{2}\left\{(M_{1,\bm r}^+\overline{\Phi_{1,\bm r}^+}+M_{1,\bm r}^-\Phi^-_{1,\bm r})\overline{V^+_{\bm r}}+{\rm h.c.}\right\}\right)\nonumber\\
&+|\phi_{1,\bm 0}^+|^2+\left(\frac{4\pi f}{g_{\rm 4D}\mathcal{A}_1}-\sum_{i=2}^3\frac{4\pi f}{g_{\rm 4D}\mathcal{A}_i}-g_{\rm 4D}|\phi_{1,\bm 0}^+|^2-g_{\rm 4D}\sum_{\bm r,i}(|\Phi_{i,\bm r}^+|^2-|\Phi_{i,\bm r}^-|^2)\right)({V}^{1}_{\bm 0}-{V}^{2}_{\bm 0}),\\
    W_{\rm 4D}=&\sum_{\bm r}\left(M^+_{3,\bm r}\Phi_{3,\bm r}^+\Phi^-_{2,\bm r}+M^-_{3,\bm r}\Phi_{3,\bm r}^-\Phi^+_{2,\bm r}\right),
\end{align}
where we have rescaled all fields by $g$. Notice that $\phi_{1,\bm 0}^+$ appears explicitly because it is not contained in the new bases. 

Let us show the (KK) mass spectrum of the charged fields. We notice the followings: $V_{\bm r}^+$ is now a complex vector superfield (in Wess-Zumino gauge), and contains two independent spinors, which have a mixed mass term with the spinors in $(\Phi^+_1,\Phi_1^-)$. The D-term potential of $V_{\bm 0}^{1,2}$ yields the scalar masses. The D-term potential of $V_{\bm r}^-$ yields the scalar mass of $\Phi_{1,\bm r}^{\pm}$. Indeed, the D-term potential derived by complete-squaring terms containing $D_{1,\bm 0},D_{2,\bm 0}$ and $D_{\bm r}^+$ (for $\bm r\neq 0$) becomes
\begin{align}
    \mathcal{L}_D=&\frac{1}{2}(D_{1,\bm 0}^2+D^2_{2,\bm 0})+|D_{\bm r}^+|^2-\frac{1}{\sqrt2}\left\{(M_{1,\bm r}^+\overline{\varphi^+_{1,\bm r}}+M_{1,r}^-\varphi_{1,\bm r}^-)\overline{D_{\bm r}^+}+{\rm h.c.}\right\}\nonumber\\
   & +\frac{1}{2}\left(\frac{4\pi f}{g_{\rm 4D}\mathcal{A}_1}-\sum_{i=2}^3\frac{4\pi f}{g_{\rm 4D}\mathcal{A}_i}-g_{\rm 4D}|\varphi_{1,\bm 0}^+|^2-g_{\rm 4D}\sum_{\bm r,i}(|\varphi_{i,\bm r}^+|^2-|\varphi_{i,\bm r}^-|^2)\right)(D_{1,\bm 0}-D_{2,\bm 0})\nonumber\\
   \underset{\text{on-shell}}{\to}&-\frac{1}{4}\left(\frac{4\pi f}{g_{\rm 4D}\mathcal{A}_1}-\sum_{i=2}^3\frac{4\pi f}{g_{\rm 4D}\mathcal{A}_i}-g_{\rm 4D}|\varphi_{1,\bm 0}^+|^2-g_{\rm 4D}\sum_{\bm r,i}(|\varphi_{i,\bm r}^+|^2-|\varphi_{i,\bm r}^-|^2)\right)^2-\frac{1}{2}\sum_{\bm r}\left|M_{1,\bm r}^+\overline{\varphi^+_{1,\bm r}}+M_{1,r}^-\varphi_{1,\bm r}^-\right|^2\nonumber\\
   =&V_{\rm flux}-\frac12m_0^2\left(|\varphi_{1,\bm 0}^+|^2+\sum_{\bm r,i}(|\varphi^+_{i,\bm r}|^2-|\varphi^-_{i,\bm r}|^2)\right)-\frac{1}{2}\sum_{\bm r}\left|M_{1,\bm r}^+\overline{\varphi^+_{1,\bm r}}+M_{1,r}^-\varphi_{1,\bm r}^-\right|^2+\cdots\nonumber\\
   =&V_{\rm flux}-\frac12m_0^2\left(|\varphi_{1,\bm 0}^+|^2+\sum_{\bm r}\sum_{i=2}^3(|\varphi^+_{i,\bm r}|^2-|\varphi^-_{i,\bm r}|^2)\right)-M_r^2|\tilde{\varphi}^1|^2+\cdots
\end{align}
where 
\begin{align}
    V_{\rm flux}\equiv-\frac{1}{4}\left(\frac{4\pi f}{g_{\rm 4D}\mathcal{A}_1}-\sum_{i=2}^3\frac{4\pi f}{g_{\rm 4D}\mathcal{A}_i}\right)^2,\label{fluxpotential}
\end{align}
$m_0^2\equiv -m_{1,1}^2+ \sum_{i=2}^3m_{i,1}^2$, ellipses denote terms of $\mathcal{O}(\varphi^3)$, and we have defined a mass eigenmode
\begin{align}
    \tilde{\varphi}^1\equiv \frac{1}{\sqrt2 M_r}\left(M_{1,\bm r}^-\overline{\varphi^+_{\bm 1,\bm r}}+M_{1,\bm r}^+\varphi_{1,\bm r}^-\right).
\end{align}
Note that another mode orthogonal to $\tilde{\varphi}^1$ corresponds to the Nambu-Goldstone mode and is eaten by the massive KK vectors, and therefore the bosons in $(V^{+}_{\bm r},\Phi_{1,\bm r})$ have the common mass $M_r$. 
Taking account of all, we find\vspace{10pt}\\
\fbox{Fermion masses}
\begin{align}
   M^2_f=\left\{\begin{array}{ll}
   0& (\text{for }\phi_{1,\bm0}^+)\\
   (M_{1,\bm r}^+)^2&(\text{for  }\Phi^+_{1,\bm r}, \lambda_{\bm r}^+)\\
   (M_{1,\bm r}^-)^2 & (\text{for }\Phi^-_{1,\bm r}, \lambda_{\bm r}^-)\\
   (M_{3,\bm r}^+)^2& (\text{for }\Phi^-_{2,\bm r}, \Phi^+_{3,\bm r})\\
   (M_{3,\bm r}^-)^2&(\text{for }\Phi^+_{2,\bm r}, \Phi^-_{3,\bm r})
   \end{array}
   \right. .\label{SSMF}
\end{align}
Here, $\lambda^\pm_{\bm r}$ denote two gauginos in $V_{\bm r}^+$.\vspace{10pt}\\
\fbox{Boson masses}
\begin{align}
    M^2_b=\left\{\begin{array}{ll}\frac12m_0^2 &(\text{for }\phi_{1,\bm 0}^+)\\ 
    M_r^2&(\text{for }V_{\bm r}^-, \Phi^\pm_{1,\bm r})\\
    (M_{3,\bm r}^+)^2+\frac12m_0^2&(\text{for }\Phi_{3,\bm r}^+)\\
    (M_{3,\bm r}^+)^2-\frac12m_0^2&(\text{for }\Phi_{2,\bm r}^-)\\
    (M_{3,\bm r}^-)^2+\frac12m_0^2&(\text{for }\Phi_{2,\bm r}^+)\\
    (M_{3,\bm r}^-)^2-\frac12m_0^2&(\text{for }\Phi_{3,\bm r}^-)\\
    \end{array}\right. .\label{SSMB}
\end{align}
We will use these results in \eqref{SSMF2} and \eqref{SSMB2}.

\subsection{``Irregular'' KK towers in the superfield model}\label{irregKK}
We explicitly show the decomposition of the KK tower which is used to compute the vacuum energy. Only nontrivial sectors are $\Phi^-_{2,\bm r}$ and $\Phi_{3,\bm r}^+$, which follow the condition~\eqref{Phi23cond}. In computing the vacuum energy, we decompose the tower as \eqref{decompose}, but \eqref{Phi23cond} removes some partial towers. We find\vspace{10pt}\\
\fbox{Fermions}
\begin{align}
    \Phi^-_{2,\bm r},\Phi_{3,\bm r}^+: &\cancel{(0,0,0)},\quad \cancel{(\tilde{r}_1,0,0)},\quad (0,\tilde{r}_2,0),\quad (0,0,\tilde{r}_3)\nonumber\\
    &(\tilde{r}_1,\tilde{r}_2,0),\quad (\tilde{r}_1,0,\tilde{r}_3),\quad (0,\tilde{r}_2,\tilde{r}_3),\quad (\tilde{r}_1,\tilde{r}_2,\tilde{r}_3),
    \end{align}
    \fbox{Bosons}
    \begin{align}
    \Phi^-_{2,\bm r}: &\cancel{\left(\frac12,-\frac12,-\frac12\right)},\ \cancel{\left(\tilde{r}_1+\frac12,-\frac12,-\frac12\right)},\ \left(\frac12,\tilde{r}_2-\frac12,-\frac12\right),\ \left(\frac12,-\frac12,\tilde{r}_3-\frac12\right)\nonumber\\
    &\left(\tilde{r}_1+\frac12,\tilde{r}_2-\frac12,-\frac12\right),\ \left(\tilde{r}_1+\frac12,-\frac12,\tilde{r}_3-\frac12\right),\ \left(\frac12,\tilde{r}_2-\frac12,\tilde{r}_3-\frac12\right), \left(\tilde{r}_1+\frac12,\tilde{r}_2-\frac12,\tilde{r}_3-\frac12\right),\\
    \Phi^+_{3,\bm r}: &\cancel{\left(-\frac12,\frac12,\frac12\right)},\ \cancel{\left(\tilde{r}_1-\frac12,\frac12,\frac12\right)},\ \left(-\frac12,\tilde{r}_2+\frac12,\frac12\right),\ \left(\frac12,\frac12,\tilde{r}_3+\frac12\right)\nonumber\\
    &\left(\tilde{r}_1-\frac12,\tilde{r}_2+\frac12,\frac12\right),\ \left(\tilde{r}_1-\frac12,\frac12,\tilde{r}_3+\frac12\right),\ \left(-\frac12,\tilde{r}_2+\frac12,\tilde{r}_3+\frac12\right), \left(\tilde{r}_1-\frac12,\tilde{r}_2+\frac12,\tilde{r}_3+\frac12\right),
    \end{align}
    where we have used the vector notation $(r_1,r_2,r_3)$ and $\tilde{r}_i\in {\mathbb N}$.
    
\section{Comment on dimensional reduction in magnetized tori models}\label{dimred}
We discuss a feature of the dimensional reduction within the magnetized tori models. In the case without magnetic fluxes, we may consider the dimensional reduction e.g. from ${\mathbb R}^{1,3}\times ({\mathbb T}^3)$ to ${\mathbb R}^{1,3}\times ({\mathbb T}^2)$ simply by taking the radius of a torus to be zero, which makes corresponding KK modes infinitely heavy. Then, such KK modes decouple from theory and we are left with the KK modes only of the remaining ${\mathbb T}^2$. However, in the presence of magnetic fluxes, such a naive dimensional reduction is inconsistent due to the possible tachyonic modes from $W_{z(\bar z)}^{ab}$-sector. If one takes the limit $\mathcal{A}_n\to 0$ for any torus, we would find infinitely tachyonic modes. This fact would be related to the topological nature of the magnetic flux, and that theories with different numbers of magnetic fluxes are disconnected at least perturbatively. Therefore, we are not able to perform a simple dimensional reduction. Note that, one may think that tachyonic mode appears from the $W_{z(\bar z)}^{ab}$ of the vanishing torus direction and the tachyonic modes should be truncated first. It is true but yet we cannot take a naive dimensional reduction as the mass of $W_{z(\bar z)}^{ab}$-sector depends on the vanishing torus area. Even after truncating the tachyonic modes, $\mathcal{A}_n\to 0$ leads to the decoupling limit of all the bosonic KK modes. It is worth noting that for spin 1/2 fields, the naive dimensional reduction works, and the limit $\mathcal{A}_n\to 0$ leaves the KK torwer in a lower dimension. 

From these considerations, in the magnetized tori models, one needs to consider the models of different dimensions separately. In this work, we will only discuss the 10D SYM case, but the computational techniques used here would be applicable to 8D models.\footnote{For 6D models, tachyonic modes from vector multiplet would be inevitable. In order to construct realistic models, we would need to consider at least 8D models. (It is not necessarily the case if the gauge sector is just Abelian group and charged matters are only from hypermultiplets.) }

\end{document}